\documentclass[12pt]{article}
\usepackage{amsfonts}
\usepackage{amssymb}
\usepackage{amsmath}
\usepackage{amssymb}
\usepackage{amsfonts}
\usepackage{listings}
\usepackage{epsfig}
\usepackage{subfigure}
\usepackage{gensymb}
\usepackage{caption}
\usepackage{graphicx}
\usepackage{authblk}
\usepackage{multirow}
\usepackage{tabularx}
\usepackage[breaklinks,colorlinks,linkcolor=blue,citecolor=blue]{hyperref}
\usepackage[all]{hypcap}

\date{}       
\begin{document}
	
\title{Volatility models applied to geophysics and high frequency financial market data}

\author{ Maria C. Mariani\footnote{Department of Mathematical Sciences and Computational Science Program, University of Texas at El Paso.}, Md Al Masum Bhuiyan\footnote{Computational Science Program, University of Texas at El Paso.}, Osei K. Tweneboah\footnote{Computational Science Program, University of Texas at El Paso} \\Hector Gonzalez-Huizar\footnote{Department of Geological Sciences, University of Texas at El Paso.}, and Ionut Florescu\footnote{School of Business, Stevens Institute of Technology, Hoboken.}}



\maketitle

\begin{abstract}
This work is devoted to the study of modeling geophysical and financial time series. A class of volatility models with time-varying parameters is presented to forecast the volatility of time series in a stationary environment. The modeling of stationary time series with consistent properties facilitates prediction with much certainty. Using the GARCH model for financial stationary data and the stochastic volatility model for geophysical stationary data, we forecast one-step-ahead suggested volatility with $\pm 2$ standard prediction errors, which is enacted via Maximum Likelihood Estimation. We compare the stochastic volatility model relying on the filtering technique as used in the conditional volatility with the GARCH model. We conclude that the stochastic volatility is a better forecasting tool than GARCH (1,1), since it is less conditioned by autoregressive past information. 
\end{abstract}
	
\begin{keywords}
ADF test; KPSS test; Financial time series; Geophysical time series; GARCH model; Maximum likelihood estimation;  Stochastic volatility model; Seismogram
\end{keywords}
\newpage
\section{Introduction}
\label{sec1}
Forecasting of time series with estimation of time-varying parameters is very important in modeling the dynamic evolution of the volatility. It is assumed that a model that attracts the attention of investors can potentially be used to predict key variables, for instance, returns, volatility, and volume of stock market. It is to be noted that the development of forecasting methodologies in geophysics helps us to identify the type of source that generates a recorded seismic signal. This type of methodologies is generally applied to various fields, such as finance, geophysics, and safety of power system \cite{Fong}. So, a reliable technique of forecasting, including the related time information, is essential to construct less risky portfolios or to make higher profits.

Financial time series manifests typical non-linear characteristics, and they involve volatility clustering where the returns indicate their dynamism. In this study, we develop a volatility forecasting method in which the logarithm of the conditional volatility follows an autoregressive time series model. R.F. Engle's paper \cite{Engle} introduced the autoregressive conditional heteroskedasticity (ARCH) model to express the conditional variance of available returns as a function of previous observations. Few years later, S. Bollerslev \cite{bol} modified this concept and generalized the ARCH (GARCH) model that allows the conditional variance to depend on the previous conditional variance as well as on the squares of previous returns. In other words, the system volatility in GARCH model is driven by the observed values in a pre-deterministic fashion. In fact, over the past few decades, a considerable amount of deterministic models has been suggested to forecast the volatility in finance. The reason is that they are very simple, and help to account for clustered errors and nonlinearity issues. In the present study, we first propose a continuous-time stationary and GARCH (1,1) process that is useful in the analysis of high frequency financial time series. 

It is now widely believed that the measurements of a sequence of geophysics and finance are stochastically dependent on the time needed. In other words, there is a correlation among the numbers of data points at successive time intervals. In Ref. \cite{Mariani} and \cite{Mariani2} , the authors used stochastic models to describe a unique type of measurement dependence in geophysical and financial data. It has been observed that the data may follow different behaviors over time, for instance, the mean reversion and fluctuation of power spectrum. Such observations are unlike those of the classical modeling foundations. But the concept of time-dependent observations suggests that the current information needs to be evaluated on the basis of its past behavior \cite{Hamiel}. This behavior of time series makes it possible to effectively forecast volatility and to obtain some stylized facts, namely, time-varying volatility, persistence, and clustering. 

A distinctive feature of the time series is that the deterministic model i.e. GARCH does not allow for a full statistical description of volatility \cite{Brockman}. When there are high fluctuations in the time series, the GARCH (1,1) model predicts the volatility arbitrarily since it cannot capture the high volatile nature of the data. So in this work, we propose a stochastic model and filtering technique as a way to estimate parameters. We therefore study a continuous-time stationary sequences corresponding to seismograms of mining explosions and small intraplate earthquakes to forecast the volatility by using estimated parameters. These stationary sequences are very effective to capture the characteristic of time-varying parameters in an appropriate way \cite{Mariani1}. We determined the adequacy and stationarity of the data by computing the estimated standard error and some powerful tests respectively, which will be discussed in Sections \ref{sec4} and \ref{sec5} of this paper. The main difficulty of SV model is to fit it into the data (with higher accuracy in a stochastic process), since their likelihood estimations involve numerical integration over higher dimensional intractable integrals \cite{Rubio}, whose maximization is rather complex. 

The paper is organized as follows: Section \hyperref[sec2]{2} describes the methodology of forecasting the volatility of time series. The techniques for estimating the model parameters will also discussed. Section \hyperref[sec3]{3} is devoted to the background of data. We discuss some essential parts of the financial and geophysical data properties. In section \hyperref[sec4]{4} we perform tests that analyzes the stationarity of the time series. Sections \hyperref[sec5]{5} and \hyperref[sec6]{6} provide the results when our models are applied to the data sets. These sections also study the suitability of our model with reference to the estimation of model parameters and the prediction of the volatility of time series. Finally, section \hyperref[sec7]{7} contains the conclusion and makes a comparison between the models. 
\section{Methodology}
\label{sec2}
This section describes the volatility models that will be used to forecast the time series regarding finance and geophysics. We will discuss some techniques that will be convenient to estimate the parameters of the proposed models. 
\subsection{Filtering Approach}
\label{subsec1}
The state space model is defined by a relation between the $m$-dimensional observed time series, ${\bf y}_t$, and the $n$-dimensional state vector (possibly unobserved), ${\bf x}_t$ \cite{Commandeur}. An observed equation is driven by the stochastic process as follows: 
\begin{equation}
	\label{eq:1}
	{\bf y}_t={\bf Z}_t{\bf x}_t +{\bf \epsilon}_t,
\end{equation}
where ${\bf Z}_t$ is a $m \times n$ observation matrix, ${\bf x}_t$ is a vector of $n \times 1$, and ${\bf \epsilon}_t$ is a Gaussian error term (${\bf \epsilon}_t \sim N(0,{\bf \beta}_t)).$\\
The unobservable vector ${\bf x}_t$  is generated from the transition equation which is defined as:
\begin{equation}
	\label{eq:2}
	{\bf x}_t={\bf T x}_{t-1} +{\bf \delta}_t,
\end{equation}
where ${\bf T}$ is a $n \times n$ transition matrix and  ${\bf \delta}_t \sim$ i.i.d $N(0,{\bf \zeta_t)}$. We assume that the process starts with a Normal vector $\bf {x_0}$. From Eqs. (\ref{eq:1}) and (\ref{eq:2}), we make estimation for the underlying unobserved data ${\bf x}_t$ from the given data ${\bf Y}_m = \{y_1,\ldots, y_m\}$. When $m=t$, the process is called filtering.

\subsection{Likelihood Approximation}
\label{subsec2}
Let $\varphi$ denote the parameters of the state space model, which are embedded in the system matrices ${\bf Z}_t, \bf{T},{\bf \beta}_t,$  and ${\bf \zeta}_t$. These parameters are typically unknown, but estimated from the data $Y = {y_1,\ldots,y_m}$.

The likelihood $L(\varphi \big|Y)$ is a function that assigns a value to each point in the parameter space $\Delta$ which suggests the likelihood of each value in generating the data. However, the likelihood is proportional to the joint probability distribution of the data as a function of the unknown parameters. The maximum likelihood estimation means the estimation of the value of $\varphi \in \Delta$ that is most likely to generate the vector of the observed data $y_t$ \cite{Eliason}. We may represent this as:
\begin{equation}
	\label{eq:3}
	\hat{\varphi}_{MLE} = \smash{\displaystyle\max_{\varphi \in \Delta}}L(\varphi \big|Y)=\smash{\displaystyle\max_{\varphi \in \Delta}}L_Y(\varphi)=\smash{\displaystyle\max _{\varphi \in \Delta}} \prod_{t=1}^{m}{f(y_t\big|y_{t-1}; \varphi)},
\end{equation}

where $\hat{\varphi}$ is the maximum likelihood estimator of ${\varphi}$. Since the natural logarithm function increases on $(0,\infty)$, the maximum value of the likelihood function, if it exists, occurs at the same points as the maximum value of the logarithm of the likelihood function. In this paper, we propose to work with the log-likelihood function which is defined as:
\begin{equation}
	\label{eq:4}
	\hat{\varphi}_{MLE} = \smash{\displaystyle\max_{\varphi \in \Delta}}lnL(\varphi \big|Y)=\smash{\displaystyle\max_{\varphi \in \Delta}}lnL_Y(\varphi)=\smash{\displaystyle\max _{\varphi \in \Delta}} \sum_{t=1}^{m}ln{f(y_t\big|y_{t-1}; \varphi)}.
\end{equation}

Since this is a highly non-linear and complicated function of the unknown parameters, we first consider the initial state vector ${\bf x}_0$ and develop a set of recursions for the log-likelihood function with its first two derivatives \cite{Gupta}. We then use Newton-Raphson algorithm \cite{Jones} successively until the negative of the log-likelihood is minimized to obtain the MLE.
\subsection{GARCH model}
\label{subsec3}
The Generalized Autoregressive Conditional Heteroscedasticity (GARCH) model \cite{Engle, bol}, was introduced in order to model the fluctuations of the variances of financial data. It is conditional, because in these models, the nature of subsequent volatility is conditioned by the information of the current period. Heteroscedasticity refers to non-constant volatility. The observations $y_t$ of high frequency financial time series used in this paper may be represented as: 
\begin{align}
	\label{eq:5}
	y_t  & =  \sigma_t \eta_t,  
\end{align}
where $\sigma_t$ is the volatility of the observations and $\{\eta_t\}_{t\in \mathbb{N}}$ is a Gaussian white noise sequence, independent of $\{\sigma_t\}_{t\in \mathbb{N}}$ and $\{y_t\}_{t\in \mathbb{N}}$.  This equation can be interpreted as the observation equation of a state space model (see subsection  \ref{subsec1}), whereby the state equation is a recursive formula for the state $\sigma_t$: 
\begin{align}
	\label{eq:6}
	\sigma^2_t &=a_0+a_1y^2_{t-1}+b_1\sigma^2_{t-1},
\end{align}
where  $a_0, a_1, b_1\geq0$, so that $\sigma^2_t > 0$ for any values of $y_t$. Eqs. (\ref{eq:5}) and (\ref{eq:6}) admit a non-Gaussian ARMA (1,1) model \cite{Rajarshi} for the squared process as: 
\begin{equation}
	\label{eq:7}
	y^2_t=a_0+(a_1+b_1)y^2_{t-1}+\phi_t-b_1\phi_{t-1},
\end{equation}
where $\phi_t=\sigma_t^2(\eta_t^2-1)$. In order to compute the variance at time t, we follow the standard GARCH $(m, n)$ model which is of the form: 
\begin{equation}
	\label{eq:8}
	\sigma^2_t =a_0 +\sum_{j=1}^{m}a_jy_{t-j}^2 + \sum_{j=1}^{n}b_j\sigma_{t-j}^2.
\end{equation}
If $n=0$ then the GARCH model changes into  an ARCH (m) model.

The parameters $a_0, a_i,$ and $b_j$ are estimated by MLE (subsection \ref{subsec2}) using the lilkelihood function. Taking into account the Normal probability density function, the conditional likelihood in Eq. (\ref{eq:4}) is obtained from the product of Normal ($N(0,\sigma_t^2)$) densities with $\sigma_t^2$. Using  the estimated parameters, we obtain one-step-ahead prediction of the volatility ($\widehat {\sigma}^2_t$), that is, 
\begin{equation}
	\label{eq:9}
	\widehat {\sigma}^2_t =\widehat{a}_0 +\sum_{j=1}^{m}\widehat{a}_jy_{t+1-j}^2 + \sum_{j=1}^{n}\widehat{b_j}\widehat{\sigma}_{t+1-j}^2.
\end{equation}
We can analyze the residuals and squared residuals to test the Normality using some statistical tests, for instance, Jarqua-Bera test \cite{Brys}, Shapiro-Wilk test \cite{Shapiro}, Ljung-Box test \cite{Ljung}, and LM-Arch test \cite{Wang}.

\subsection{Stochastic Volatility  model}
\label{subsec4}
The stochastic volatility (SV) model implies that the volatility is driven by an innovation sequence, that is, independent of the observations \cite{janssen}. It causes volatility through an unobservable process that allows volatilities to vary stochastically. To develop the SV model, we use the log-squared observations of the time series in Eq. (\ref{eq:5}): 
$$logy^2_t = log\sigma^2_t + log\eta^2_t$$
\begin{equation}
	\label{eq:10}
	\Rightarrow g_t = v_t + log\eta^2_t, 
\end{equation}
where $g_t=logy^2_t$ and $v_t=log\sigma^2_t$. This equation is considered as the observation equation, and the stochastic variance $v_t$ is known to be an unobserved state process. Considering the autoregression, the form of $v_t$ can be expressed as:
\begin{equation}
	\label{eq:11}
	v_t=\alpha_0 + \alpha_1v_{t-1} + \omega_t,
\end{equation}
where $\omega_t$ is a white Gaussian noise with variance $\sigma_\omega^2$. Eqs. (\ref{eq:10}) and (\ref{eq:11}) constitute the stochastic volatility model by Taylor \cite{taylor1}. In this study, our approach is to estimate the parameters $\alpha_0, \alpha_1, \sigma_\omega$ and then forecast the future observations $y_{n+m}$ from $n$ data points.

To compute the observation noise, we consider the mixtures of two Normal distributions with one centered at zero. Thus we have: 
\begin{equation}
\label{eq:12}
g_t=\lambda+ v_t + \gamma_t,
\end{equation}
where $\lambda$ is the mean of log-squared observations and $\gamma_t=Q_tz_{t0} - (Q_t-1)z_{t1}$, which fulfills the following condition:
\begin{align}
z_{t0}\sim  \text{i.i.d}   \mbox{ N} (0,\sigma^2_0) \nonumber \\
z_{t1}\sim \text{i.i.d}   \mbox{ N} (\phi_1,\sigma^2_1) \nonumber\\
Q_t\sim \text{i.i.d}   \mbox{ Bernoulli }(p),  \nonumber
\end{align}
where $p$ is an unknown mixing probability. We therefore define the time-varying probabilities $\text{Pr} \{Q_t=0\} = p_0$ and $ \text{Pr} \{Q_t=1\} = p_1$,  where $ p_0+p_1=1 $.
\subsubsection{Parameter Estimation}
\label{subsubsec1}
In order to estimate the parameters, we use the filtering technique that is followed by three steps namely, forecasting, updating, and parameter estimation. In the first step, we forecast the unobserved state vector $s_t$ on time series observations as follows:
\begin{equation}
\label{eq:15}
v^t_{t+1}=\alpha_0+\alpha_1v^{t-1}_t+ \sum_{j=0}^{1} p_{tj}K_{tj}\eta_{tj},
\end{equation}
where the predicted state estimators $v^{t-1}_t=E(v_t|y_1,\ldots,y_{t-1})$. The corresponding error covariance matrix is defined as:
\begin{equation}
\label{eq:16}
U^t_{t+1}=\alpha^2_1U^{t-1}_t+ \sigma_\omega^2 - \sum_{j=0}^{1} p_{tj}K^2_{tj}\sum\nolimits_{tj}.
\end{equation}
At this point, the innovation covariances are given as $\sum_{t0}=U_t^{t-1}+\sigma_0^2$ and $\sum_{t1}=U_t^{t-1}+\sigma_1^2$, where $U^{t-1}_{t}= T U^{t-1}_{t-1} T ^t + \sigma_\omega^2$, $U^0_0 = \sum_0$, and $\sum_t=\text{var}(\eta_t)$. Furthermore, we use Kalman filter \cite{Cipra} to measure the estimates precision, which may be shown as:
\begin{equation}
\label{eq:19}
K_{t0}=\alpha_0U_t^{t-1}/(U_t^{t-1}+\sigma_0^2)
\mbox{ and }  K_{t1}=\alpha_1U_t^{t-1}/(U_t^{t-1}+\sigma_1^2).
\end{equation}
The second step deals with updating results while we have a new observation of $y_t$ at time t. The prediction errors of the likelihood function are computed using the following relations:
\begin{equation}
\label{eq:17}
\eta_{t0}=g_t-\lambda-v_t^{t-1}
\mbox{ and } \eta_{t1}=g_t-\lambda-v_t^{t-1}-\phi_1.
\end{equation}
For estimating the parameters, we complete the updating step by assessing the time-varying probabilities (for $t=1,\ldots,m$):
\begin{align*}
p_{t1} &= \frac{p_1h_1(t|t-1)}{p_0h_0(t|t-1)+p_1h_1(t|t-1)}\\
\text{and } p_{t0} &=1-p_{t1},
\end{align*}
where $h_j(t|t-1)$ is considered to be the conditional density of $y_t$, given the previous observations $y_1,\ldots,y_{t-1}$. 

Since the observation noise of this model is not fully Gaussian, it is computationally difficult to obtain the exact values of $h_j (t|t-1)$. Hence, we use a good approximation of $h_j(t|t-1)$ that provides Normal density which is: $N(v_t^{t-1}+\phi_j,\sum_{tj}),$ for $j=0,1$ and $\phi_0=0.$

Finally, we estimate the parameters ($\Theta = (\alpha_0, \alpha_1, \sigma_w$, $\sigma_0,\phi_1, \sigma_1)')$ by maximizing the expected likelihood, where the MLE is represented as:
\begin{equation}
\label{eq:21}
ln L_Y(\Theta) = \sum_{t=1}^{m}ln\big(\sum_{j=0}^{1}p_jh_j(t|t-1)\big).
\end{equation} 
\section{Dynamic behavior of the data sets}
\label{sec3}
In this section, we present the background of the time series arising in finance and geophysics. It is the dynamic behavior of the data that encourages us to apply our methodology in this paper.  
\subsection{Financial Time Series}
\label{subsec5}
\begin{figure}[!h]
 	\begin{center}
 		\includegraphics[height=3.1cm, width=14cm]{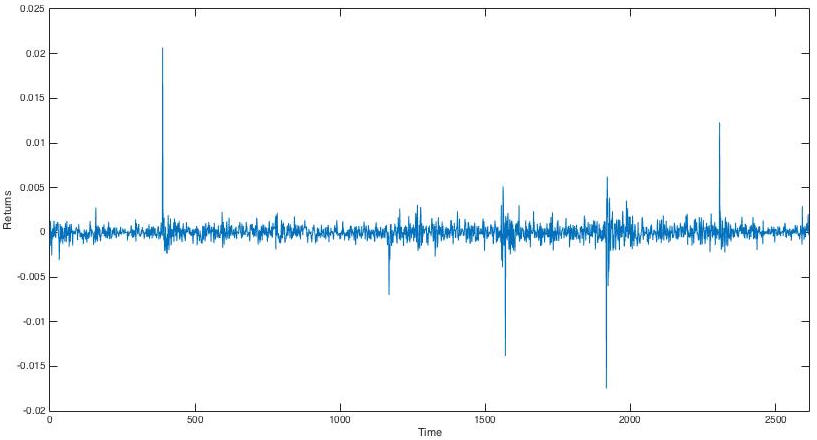}
 	\end{center}
\caption{Financial returns of high-frequency trading observations (per minute) from Bank of America stock exchange.}
 	\label{figure:fig1}
 \end{figure}

\begin{figure}[!htbp]
 	\begin{center}
 		\includegraphics[height=3.1cm, width=14cm]{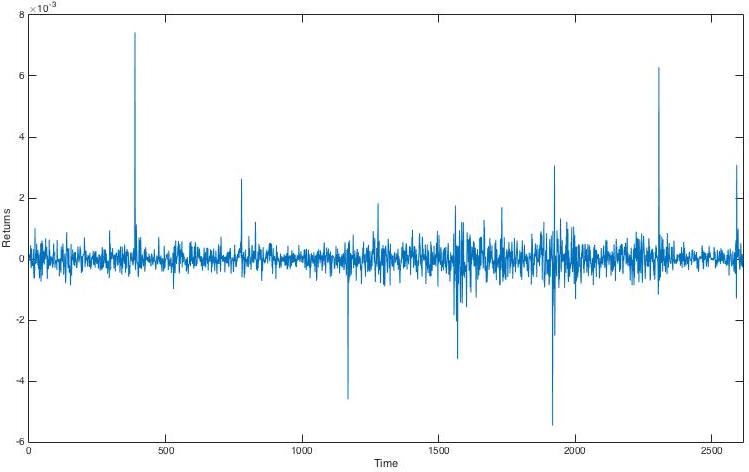}
 	\end{center}
 \caption{Financial returns of high-frequency trading observations (per minute) from  DISCOVER Financial Services stock exchange.}
 	\label{figure:fig2}
 \end{figure}

\begin{figure}[!htbp]
 	\begin{center}
 		\includegraphics[height=3.1cm, width=14cm]{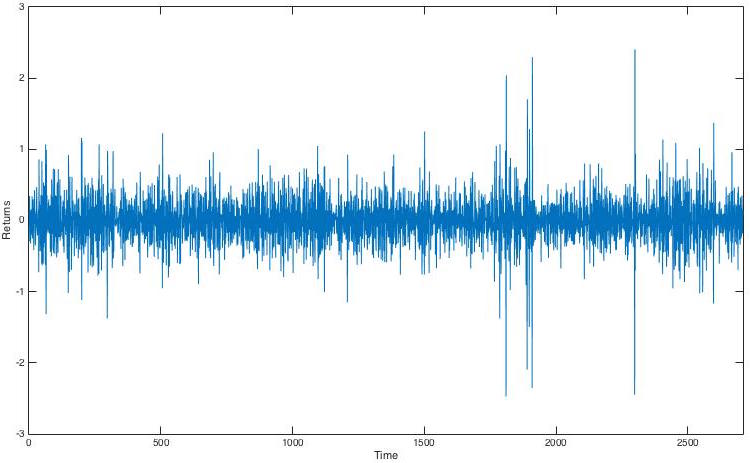}
 	\end{center}
\caption{Financial returns of high-frequency trading observations (per minute) from INTEL Corporation stock exchange.}
 	\label{figure:fig3}
 	\end{figure}

\begin{figure}[!htbp]
 	\begin{center}
 		\includegraphics[height=3.1cm, width=14cm]{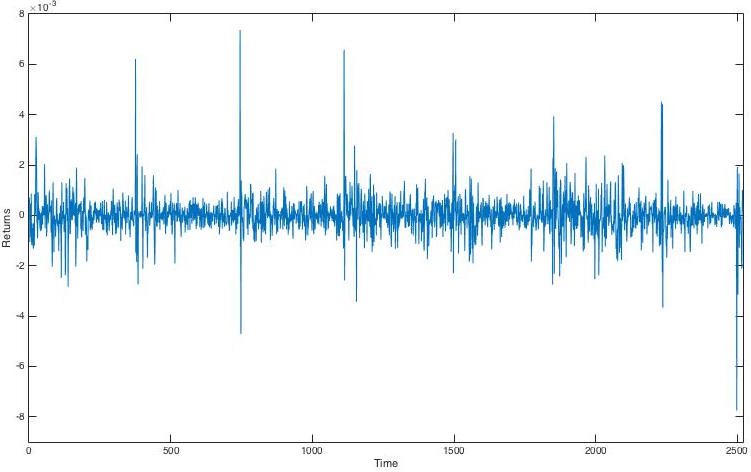}
 	\end{center}
\caption{Financial returns of high-frequency trading observations (per minute) from  IAG stock exchange.}
 	\label{figure:fig4}
 	\end{figure}

We study a set of high-frequency financial returns (per minute) from the following four stock exchanges: Bank of America Corporation (BAC), Discover Financial Services (DISCOVER), INTEL semiconductor manufacturing company (INTEL), and IAMGOLD corporation (IAG). Fig. \ref{figure:fig1}-\ref{figure:fig4} provide a good perspective on the trending direction or risk management of the high frequency returns of stock markets. The financial crisis that occurred is evident in the large spikes in the figures. We see that the volatility of data changes dramatically at a short interval and that the periods of high volatility are sometimes correlated. This establishes that the volatility itself is very volatile. The fluctuations of stock returns (per minute) typically exhibit the volatility clustering. That is to say, small changes in the price tend to be followed by small changes, and large changes by large ones. The volatility clustering suggests that current information is highly correlated with past information at different levels. 
\newpage
\subsection{Geophysical Time Series}
\label{subsec6}
The time series used in this study correspond to a set of magnitude 3.0-3.3 aftershocks of a recent magnitude 5.2 intraplate earthquake which occurred on June 26, 2014. These earthquakes were located near the town of Clifton, Arizona, where a large surface copper mine previously triggered off several explosions forming part of quarry blasts activities. We selected some explosions cataloged with similar magnitude as the earthquakes (M=3.0-3.3) and located in the same region within a radius of 10km. We collected the seismograms containing the seismic waves from two nearby seismic stations (IU.TUC and IU.ANMO) located between 150 and 400km from the seismic events. The data contains information about the date, time, longitude, latitude, the average distance to seismic events, average azimuth, and the magnitude of each seismic event in the region (see Tables \ref{table:1} and \ref{table:2}). The dynamic behavior of the geophysical time series is shown in Figs. \ref{figure:fig6}-\ref{figure:fig9}. In these figures, we notice that the frequency components change from one interval to another in earthquake or explosion as long as it lasts. The mean of the series appears to be stable with an average magnitude of approximately zero. This dynamic behavior illustrates the time evolution of the magnitude with its volatility. The volatility depends on time, in that it is high at a certain point, but low at another. The volatility clustering reflects its varying nature in time, as well as the mean reversion characteristics of the data.
\newpage
 \begin{center}
 	\captionof{table}{Stations information}
 	\begin{tabular}{ | l | l | l | l | p{3.2cm} | p{2.7cm} |}
 		\hline
 		Station  & Network & Latitude & Longitude & Average distance to events (km) & Average 
 		Azimuth (deg)\\ \hline
 		TUC & IU &  $32.3^{\circ}$ & $-110.8^{\circ}$ & 161 & 76\\ \hline
 		ANMO & IU &  $34.9^{\circ}$ & $-106.5^{\circ}$ & 357 & 224 \\ \hline
 	\end{tabular}
 	\label{table:1}
 \end{center}
 \begin{table}[!htbp]
 	\centering
 	\captionof{table}{Events information}
 	\begin{tabular}{ | l | l | l | l | l | l |}
 		\hline
 		Event  & Magnitude & Date & Time (UTC) & Latitude & Longitude\\
 		 \hline
 		Earthquake & 3.0 & 7/12/14 & 7:12:53 & $32.58^{\circ}$ & $-109.08^{\circ}$  \\
 		 \hline
 		Explosion & 3.2 & 12/23/99 & 21:15:48 & $32.65^{\circ}$ & $-109.08^{\circ}$  \\ \hline
 	\end{tabular}
 	\label{table:2}
 \end{table}
 \begin{figure}[!htbp]
 	\centering
 	\includegraphics[height=3cm, width=8cm]{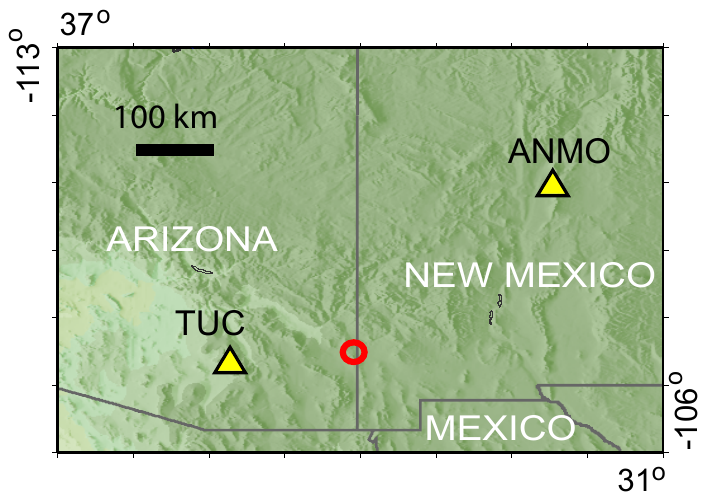}
 	\caption{The map shows the location of the seismic stations IU.TUC and IU.ANMO used in this study (yellow color triangles). Red open circle represents the area within which the earthquakes and explosions used in this study are located.}
 	\label{figure:fig5}
 \end{figure}
\begin{figure}[!htbp]
	\begin{center}
		\includegraphics[height=2.8cm, width=14cm]{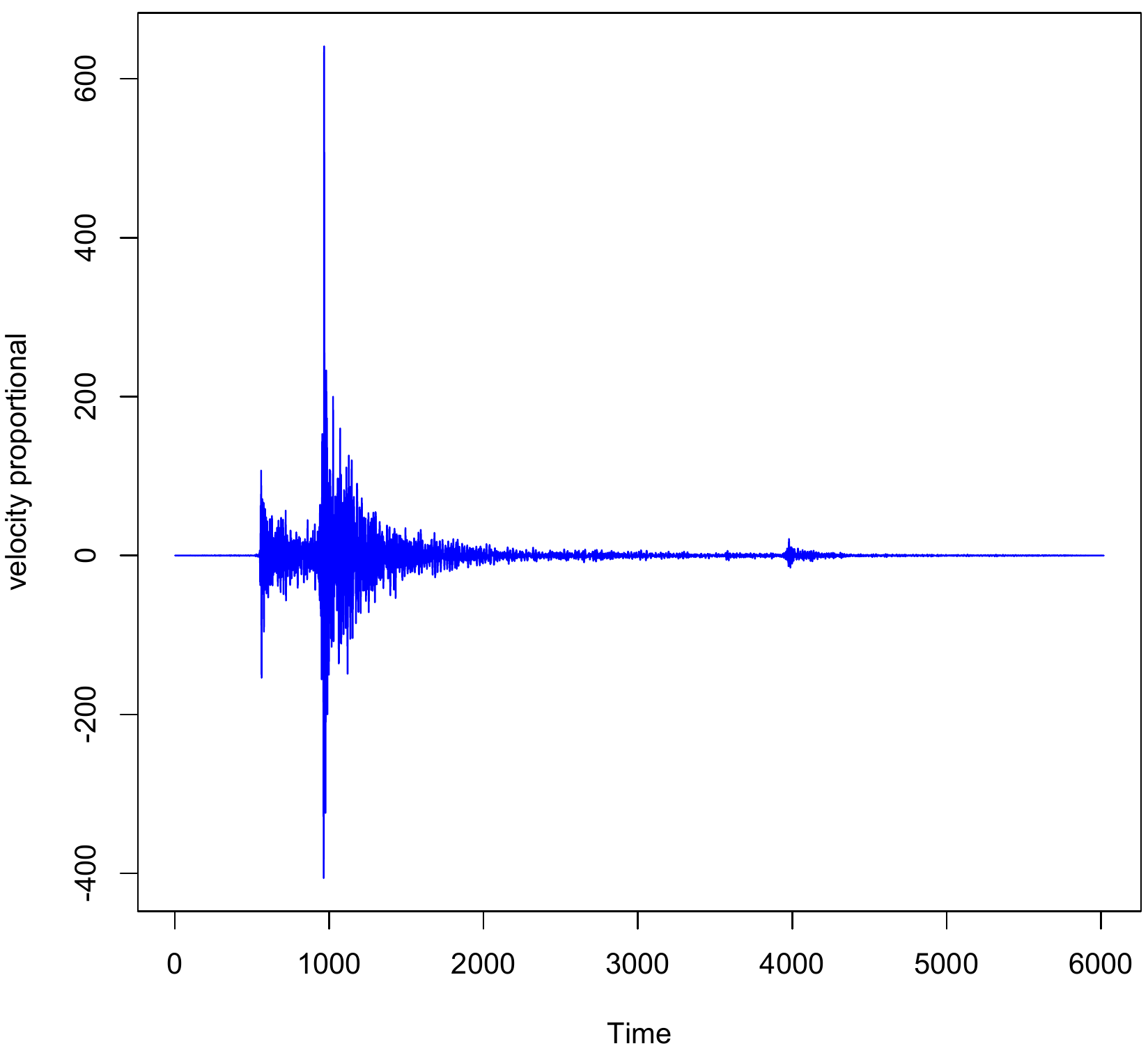}
	\end{center}
	\caption{The arrival phases from an earthquake in \hyperref[table:1]{Table-1} as recorded by TUC station.}
	\label{figure:fig6}
\end{figure}
\begin{figure}[!htbp]
	\begin{center}
		\includegraphics[height=3.3cm, width=14cm]{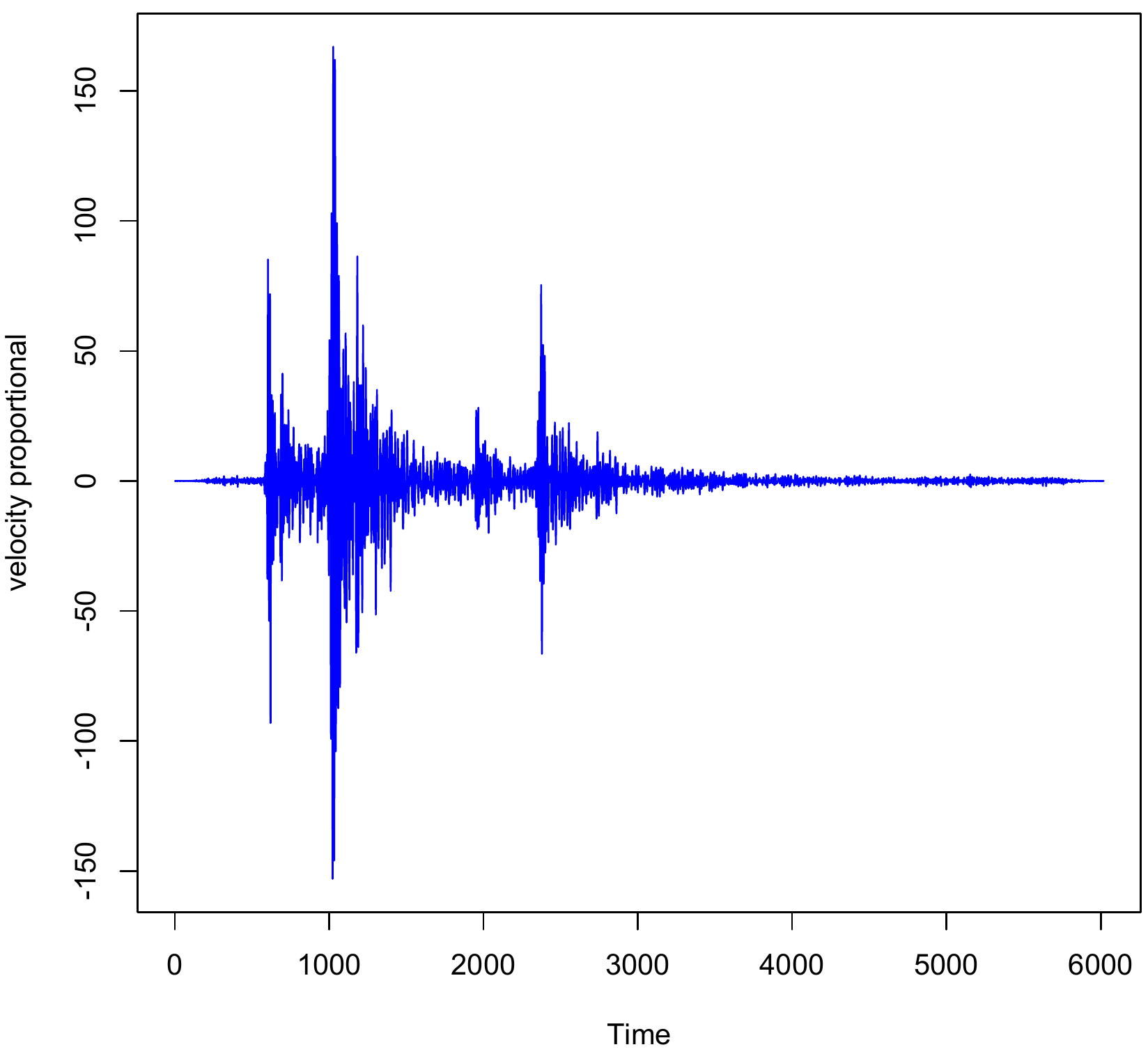}
	\end{center}
\caption{The arrival phases from an explosion in \hyperref[table:1]{Table-1} as recorded by TUC station.}
	\label{figure:fig7}
\end{figure}
\begin{figure}[!htbp]
	\begin{center}
		\includegraphics[height=3.3cm, width=14cm]{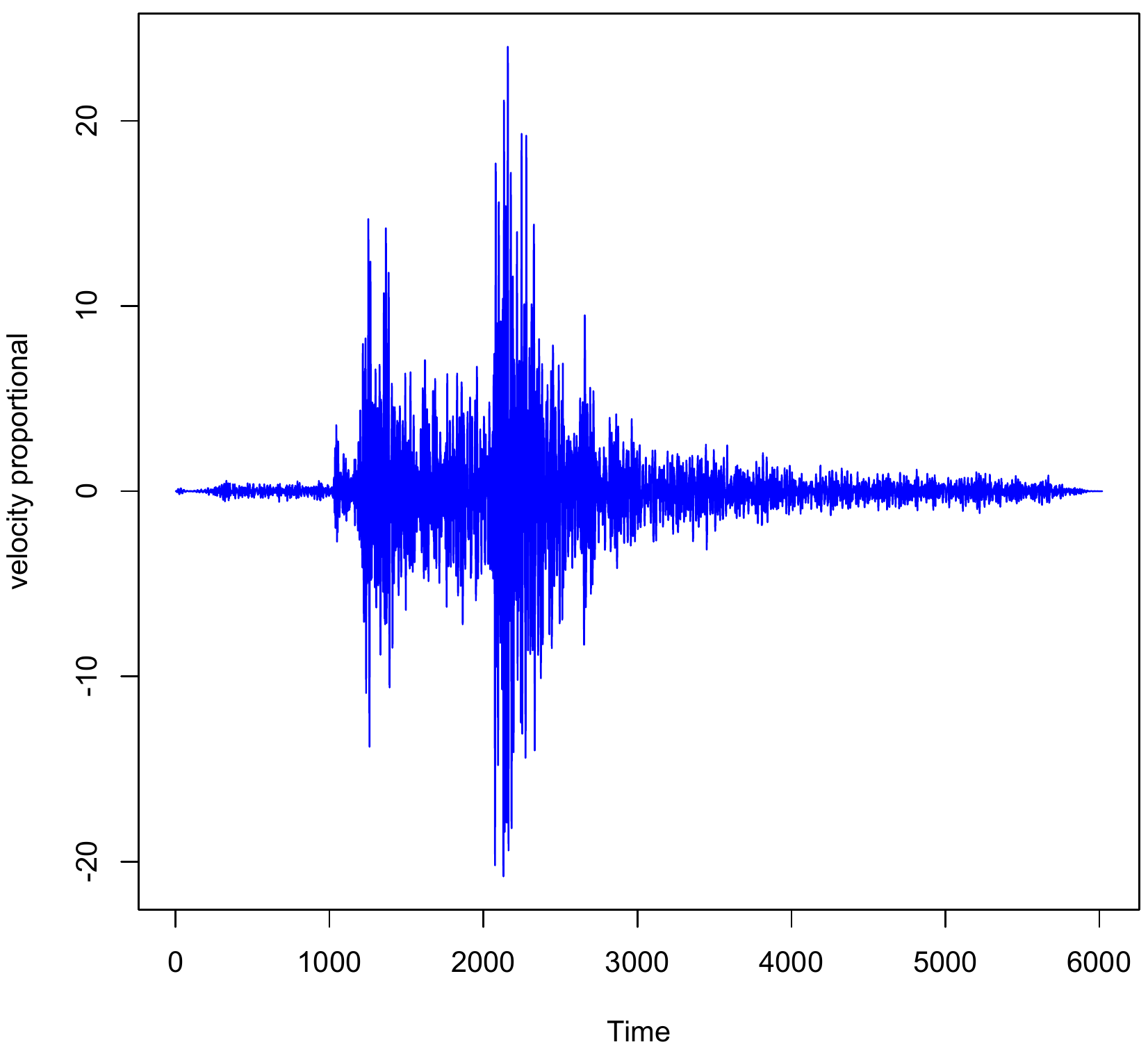}
	\end{center}
	\caption{The arrival phases from an earthquake in \hyperref[table:1]{Table-1} as recorded by ANMO station.}
	\label{figure:fig8}
\end{figure}
\begin{figure}[!htbp]
	\begin{center}
		\includegraphics[height=3.3cm, width=14cm]{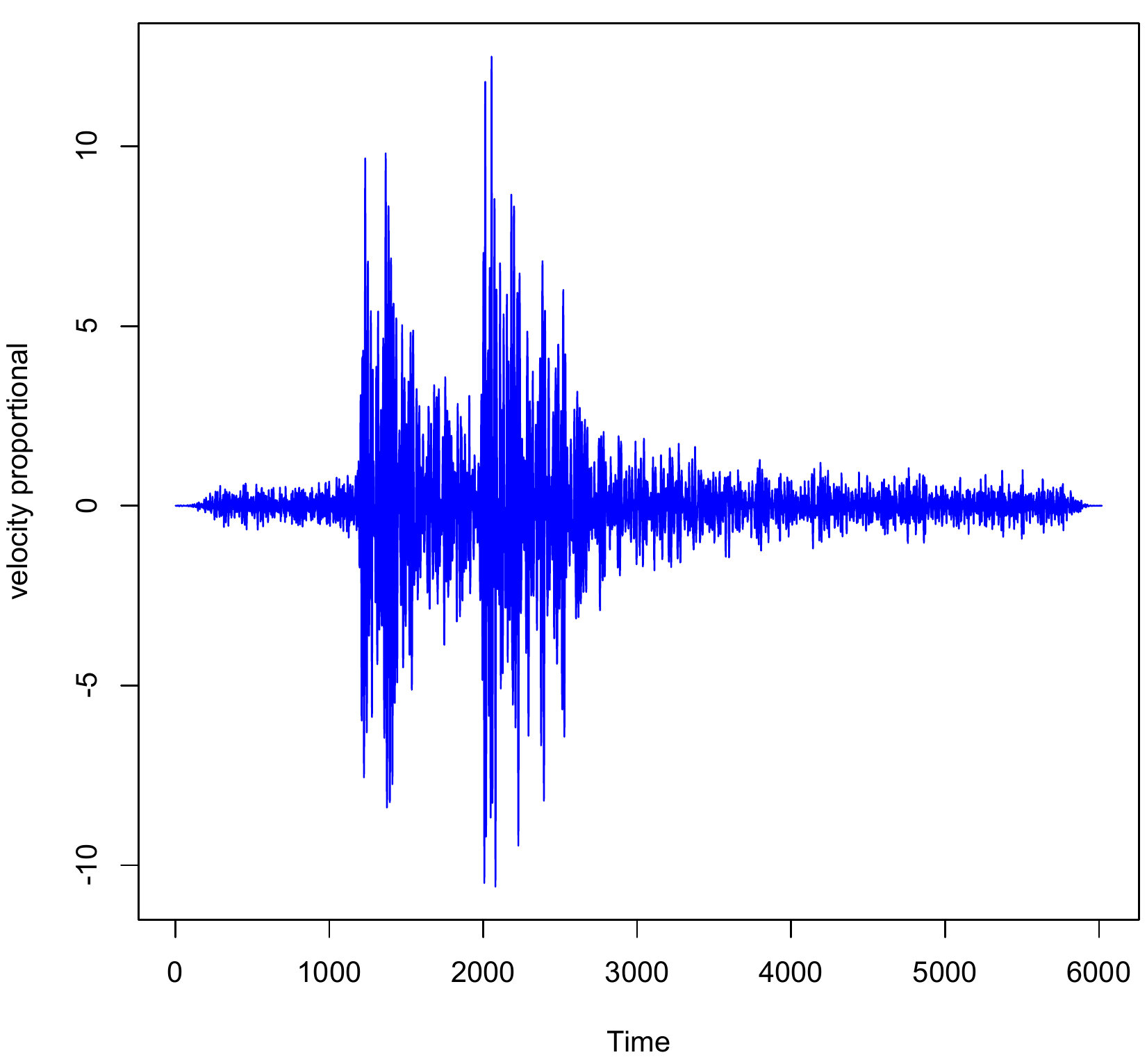}
	\end{center}
\caption{The arrival phases from an explosion in \hyperref[table:1]{Table-1} as recorded by ANMO station.}
	\label{figure:fig9}
\end{figure}
\newpage
\section{Stationary Approach}
\label{sec4}
In this section, we analyze the time series by testing for stationarity in the high frequency financial data and seismic waves generated by the earthquake and explosion data by using the Augmented Dickey Fuller (ADF) \cite{Mam} and Kwiatkowski-Phillips-Schmidt-Shin (KPSS) tests. These two tests are very powerful and capable of handling very complex models. 

\subsection{ADF test}
\label{subsec7}
The ADF test tests the null hypothesis that a time series $y_t$ is a unit root against the alternative that it is stationary, assuming that the dynamics in the data have an ARMA structure \cite{Said1984}. The summary statistics of this test for the data sets used in this work are displayed in Tables \ref{table:3} and \ref{table:4} respectively.\\
\begin{table}[!htbp]
\caption{\bf{ADF t-statistics test for financial data}} 
\centering 
\begin{tabular}{c c c c c} 
\hline\hline 
		Stocks & Test statistics & P-value \\
		\hline
		Bank of America  & -14.873 & 0.01 \\
		Discover & -13.955 & 0.01 \\
		Intel & -21.314 & 0.01\\
		IAG  & -13.830 & 0.01 \\
\hline 
\hline 
\end{tabular}
	\label{table:3} 
\end{table}
 \begin{table}[!htbp]
 	\caption{\bf{ADF t-statistics test for geophysical data}}
 	\centering 
 \begin{tabular}{c c     c  c    c c }
 	\hline \hline
 \multicolumn{1}{ c }{\multirow{2}{*}{Events} } &
  \multicolumn{2}{c}{TUC station} & \multicolumn{2}{c}{ANMO station}  \\ \cline{2-5}
 \multicolumn{1}{c }{}                        &
  Test statistics &  p-value  & Test statistics &  p-value   \\
  \hline
 		Earthquake &  -42.018 & 0.01  & -40.509 & 0.01\\ 
 		Explosion & -40.831 & 0.01  & -38.954 & 0.01\\
\hline \hline
 \end{tabular}
 \label{table:4} 
 \end{table}
 \\
 Test interpretation:\\
$H_0:$ There is a unit root for the time series.\\
$H_a:$ There is no unit root for the time series. This series is stationary.

The t-statistics are used to compute the p-values, which are compared with the significance level (0.05) and which suggest whether the null hypothesis is acceptable or not. Since the computed p-value is lower than the significance level $\alpha = 0.05$, we reject the null hypothesis $H_0$ for both financial and geophysical data, and accept the alternative hypothesis $H_a$. Thus the data under study are stationary in time. 

\subsection{KPSS test}
 \label{subsec8}
The KPSS test are used for testing a null hypothesis that an observable time series is stationary against the alternative of a unit root \cite{Kwiatkowski}. The summary statistics for the results of this test are displayed in Tables \ref{table:5} and \ref{table:6} respectively.
\begin{table}[!h]
\caption{\bf{KPSS t-statistics test for financial data}} 
\centering 
\begin{tabular}{c c c c c} 
\hline\hline 
		Stocks & Test statistics & P-value \\
			\hline
			Bank of America  & 0.1733 & 0.1 \\
			Discover & 0.1221 & 0.1 \\
			Intel & 0.0069 & 0.1\\
			IAG  & 0.3412& 0.1 \\
\hline 
\hline 
\end{tabular}
	\label{table:5} 
\end{table}
 \begin{table}[!h]
 	\caption{\bf{KPSS t-statistics test for geophysical data}}
 	\centering 
 \begin{tabular}{c c     c  c    c c }
 	\hline \hline
 \multicolumn{1}{ c }{\multirow{2}{*}{Events} } &
  \multicolumn{2}{c}{TUC station} & \multicolumn{2}{c}{ANMO station}  \\ \cline{2-5}
 \multicolumn{1}{c }{}                        &
  Test statistics &  p-value  & Test statistics &  p-value   \\
  \hline
 		Earthquake &  0.0017 & 0.1 & 0.0020 & 0.1\\ 
 		Explosion & 0.0012 & 0.1  & 0.0030& 0.1\\
\hline \hline
 \end{tabular}
 \label{table:6} 
 \end{table}
 \\
Test interpretation:\\
$H_0:$ There is a unit root for the time series. This series is stationary.\\
$H_a:$ There is no unit root for the time series.
 
As the computed p-value is greater than the significance level $\alpha = 0.05$, we accept the null hypothesis $H_0$ in all data sets. Thus the time series used in this paper are stationary time series.  

In the next two sections we describe the analysis of time series arising in geophysics and finance using a deterministic and stochastic approach. First we present the estimates when the models are applied to the data sets. To estimate the time-varying parameters, we used a deterministic and stochastic model on the time series. The idea is to compare the two techniques and observe which is suitable for forecasting the volatility. The analysis was performed by an R statistical software module. 

\section{Analysis of the Deterministic Model}
\label{sec5}
We present the results of the estimated parameters of high frequency financial data  obtained with the GARCH model. The red line in the following figures show the theoretical probability density function of Normal distribution with the same mean and standard deviation as the financial data. We therefore consider the ARCH Normality assumption on the basis of volatility $\eta_t$. 
\begin{figure}[!htbp]
    \centering
    \subfigure[BAC]
    { %
        \includegraphics[height=2.3cm,width=5cm]{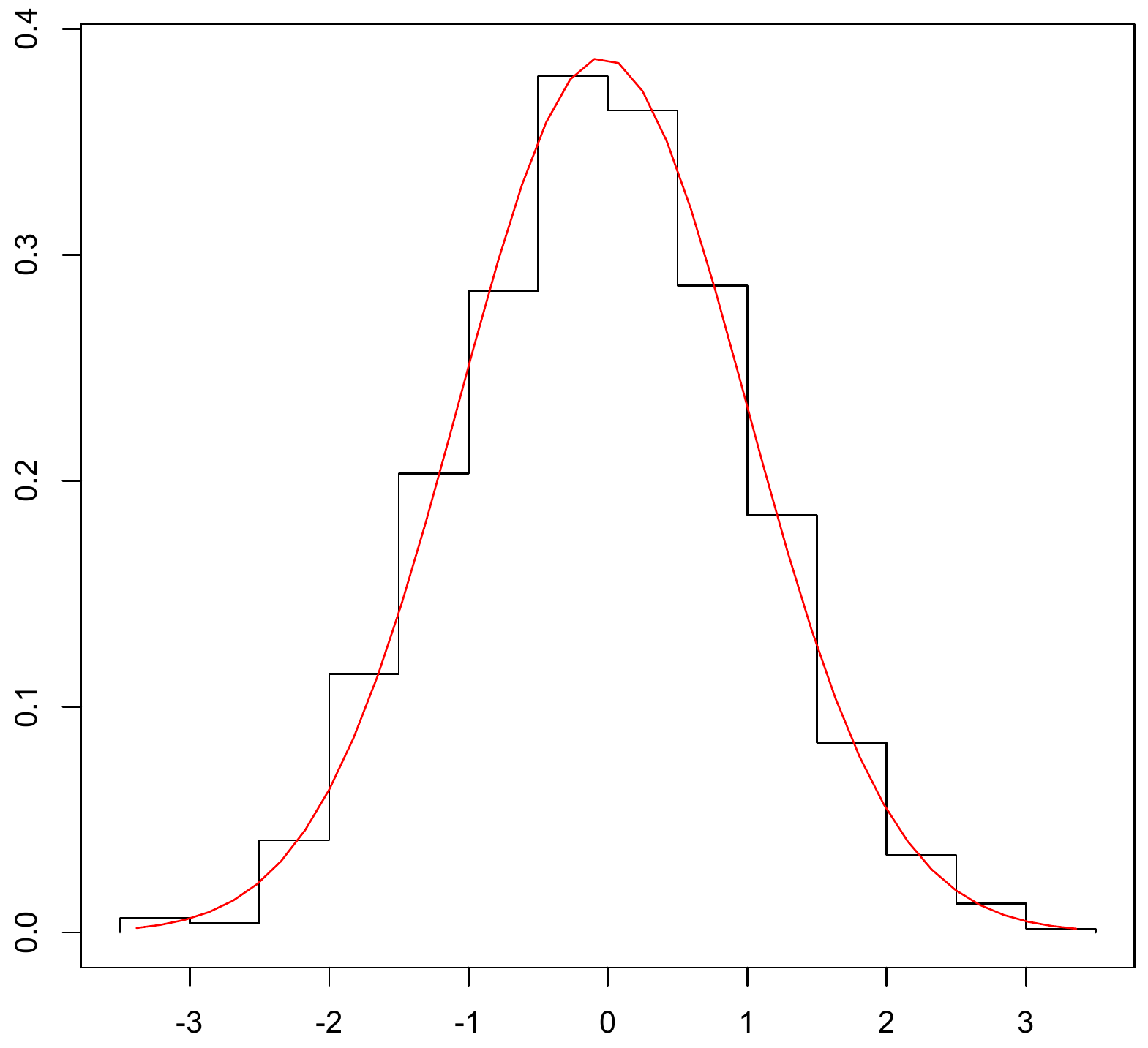}
    }
    \quad
  \subfigure[DISCOVER]
    {  
        \includegraphics[height=2.3cm,width=5cm]{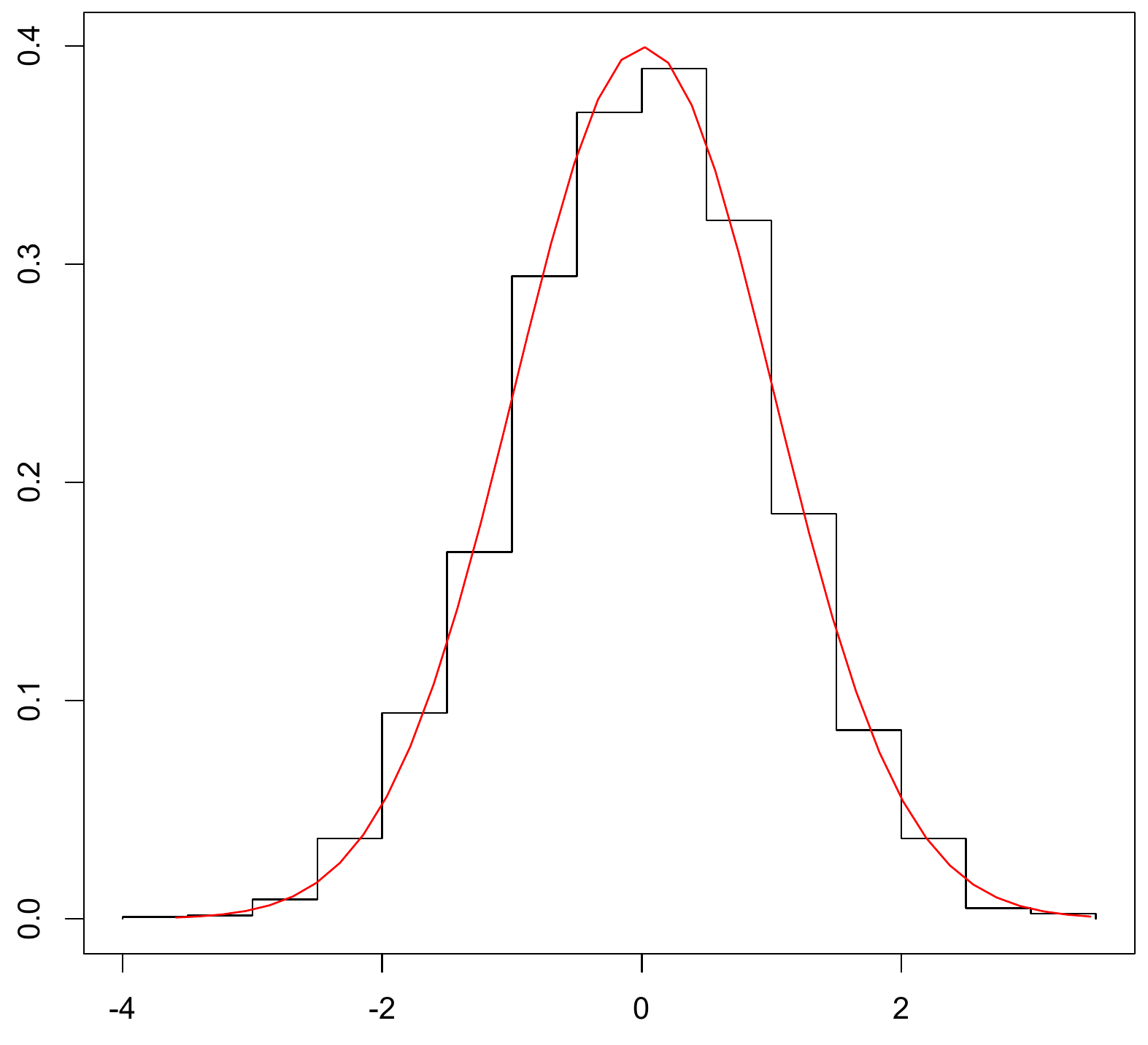}
    }
   \linebreak

    \subfigure[INTEL]
    {
        \includegraphics[height=2.3cm,width=5cm]{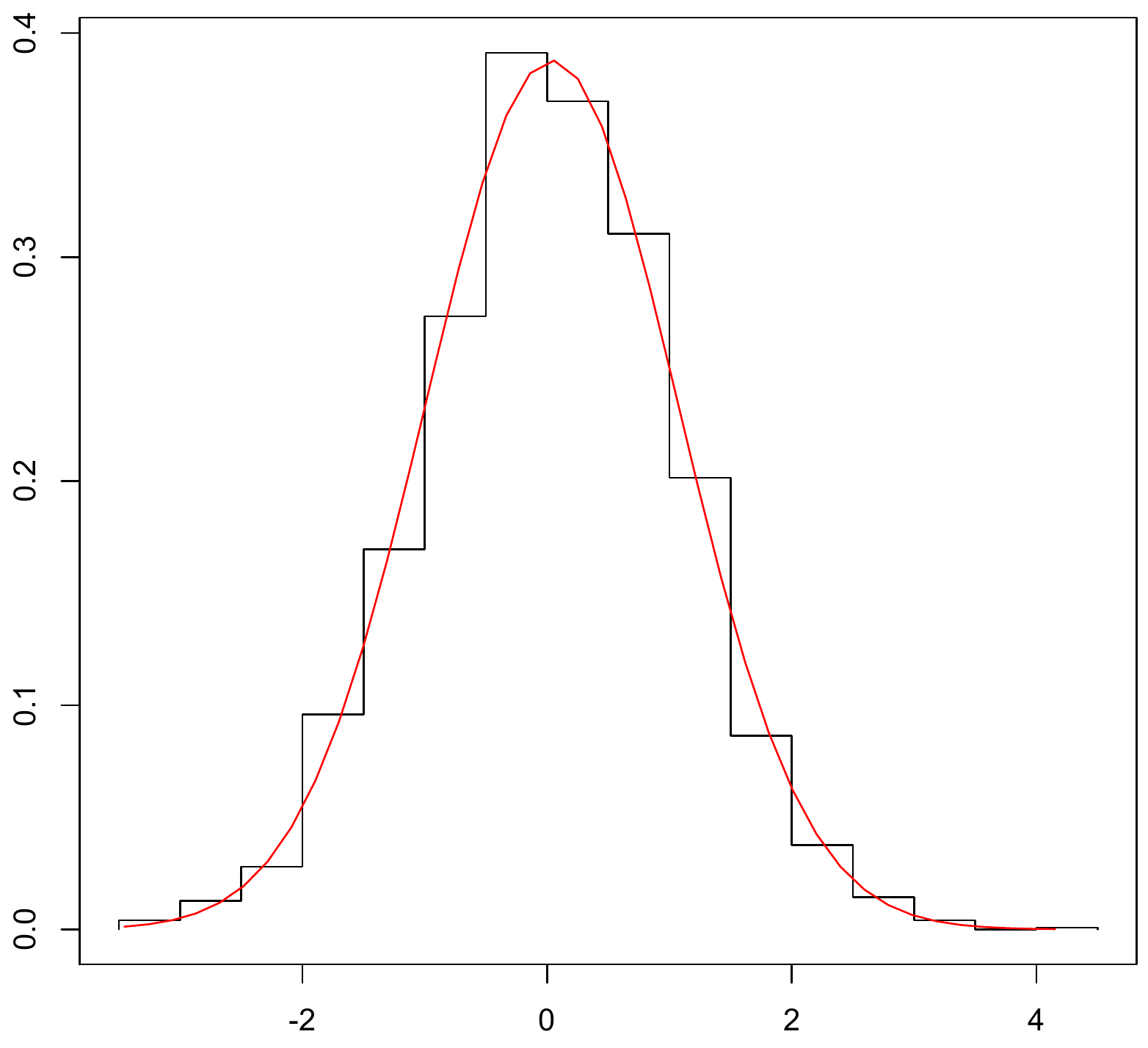}
    }
    \quad
    \subfigure[IAG]
        {
            \includegraphics[height=2.3cm,width=5cm]{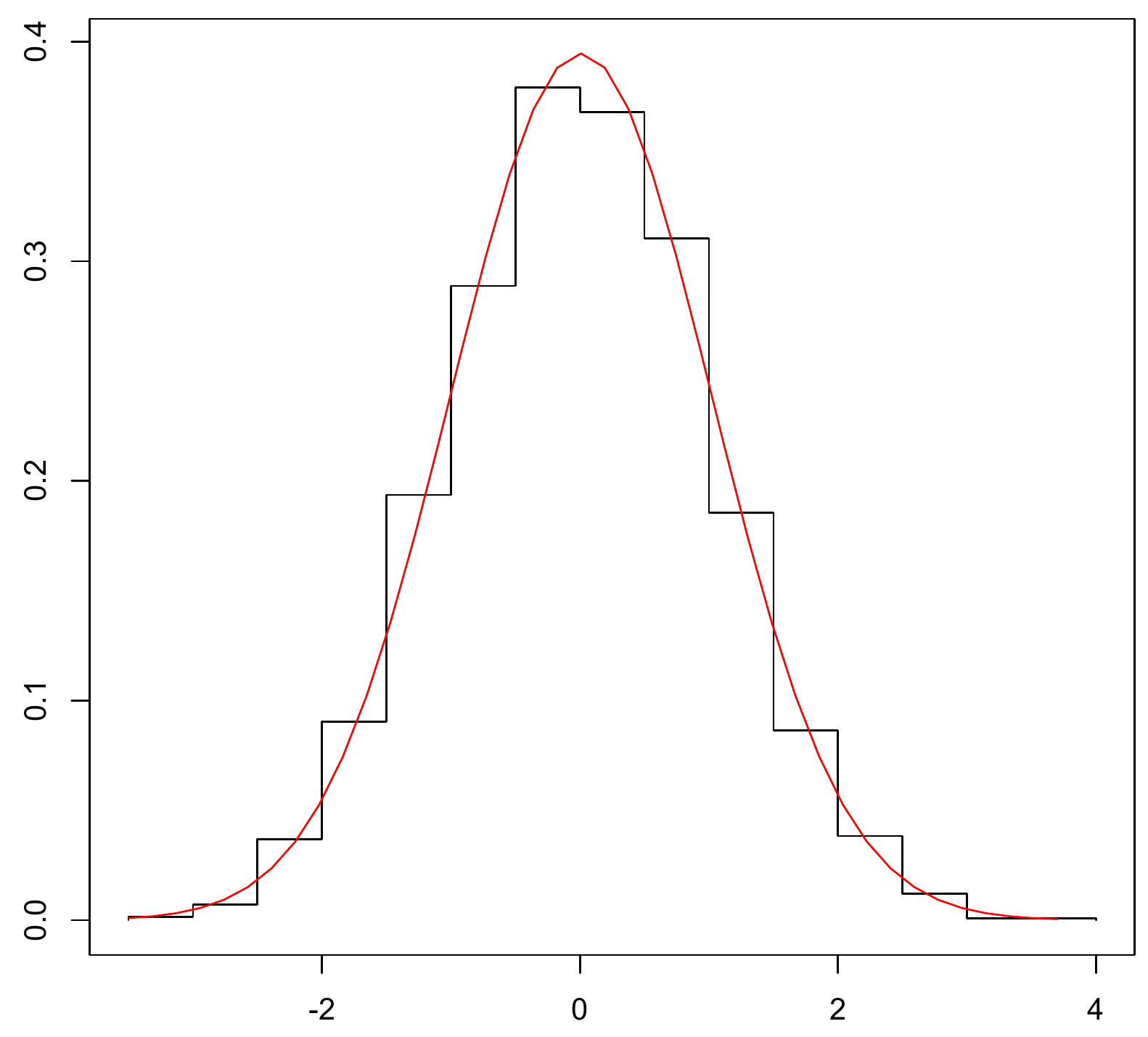}
        }
       	\caption{The histograms of financial time series and the fitted Normal density.}
          \label{figure:fig10}
\end{figure}
The estimates of the parameters $a_0, a_1$, and $b_1$ of the GARCH model are stable, as the GARCH-statistic shows ( see Tables \ref{table:7}-\ref{table:10}). Also, the estimated standard errors of the parameters in most cases are small. The smaller p-values ($<$ significance level) provide strong evidence that the GARCH $(1,1)$ model with the specified parameters is a good fit for our data. The volatility level of persistence can be determined by non-negative parameters $a_1$ and $b_1$ from these tables. We can see that the constraint $(a_1 + b_1)$ is less than 1, which is consistent with the existence of a stationary solution, and supports the results of stationary tests in section \ref{sec4}.

Tables \ref{table:11}-\ref{table:14}  summarize the standardized residuals (R) tests for  Bank of America, Discover Financial Services, INTEL semiconductor manufacturing company, and IAG stock exchanges respectively. The Jarque-Bera and Shapiro-Wilk tests of Normality strongly reject the null hypothesis that the white noise innovation process ${\eta_t}$ is Gaussian. The p-values $(>0.05)$ of Ljung-Box test for squared residuals (at lag $10, 15, 20$) and LM-Arch test suggest that the model fits the data well, with the exception of the non-normality of ${\eta_t}$. It is because the null hypothesis cannot be rejected at any reasonable level of significance.

\begin{figure}[!htbp]
\begin{center}
	{\bf {GARCH predicted volatility}}
\includegraphics[height=3.0cm, width=14cm]{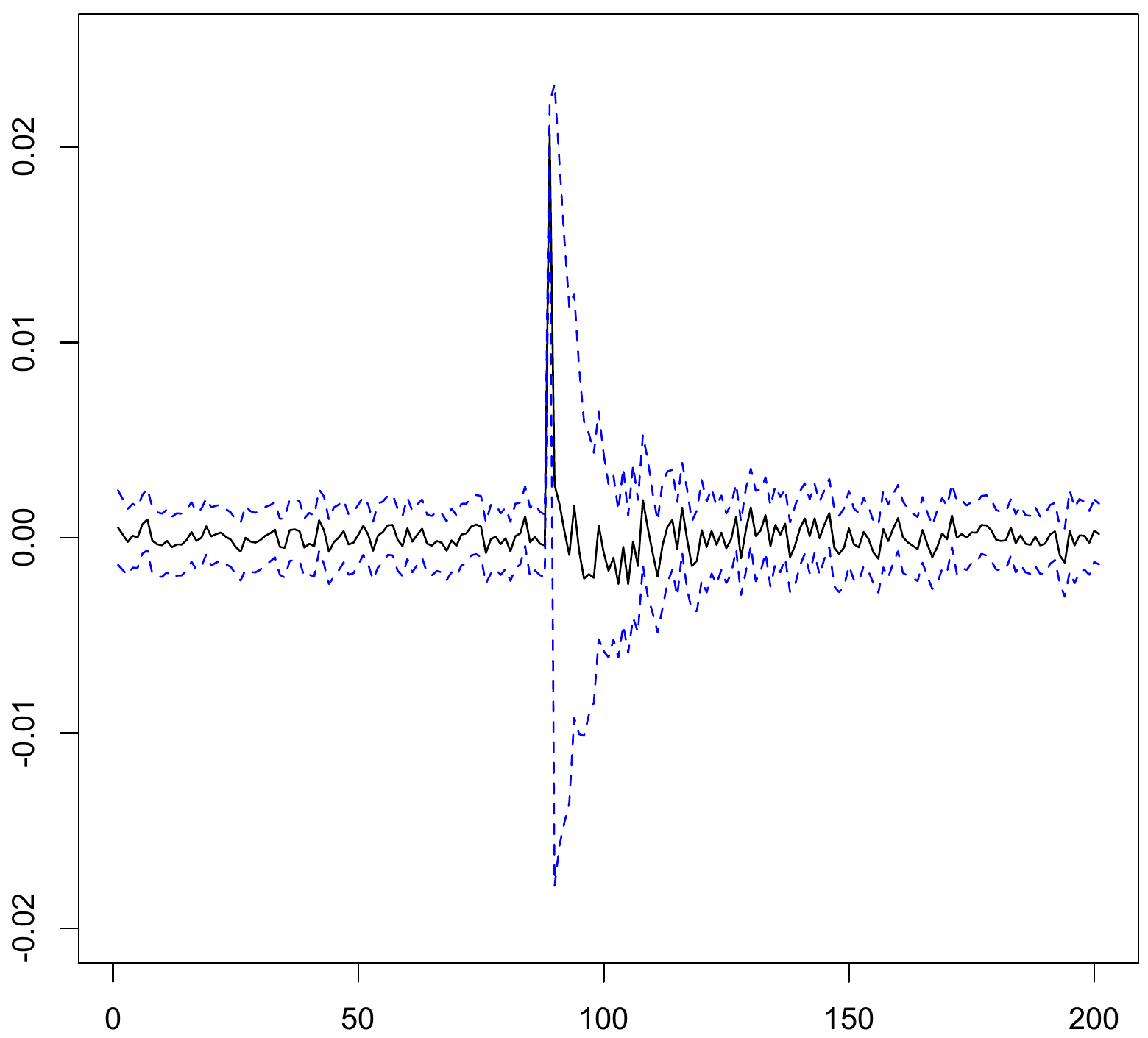}
\end{center}
\caption{One-step-ahead predicted log-volatility, with $\pm 2$ standard prediction errors for two hundred observations from BAC stock exchange.}
\label{figure:fig11}
\end{figure}

\begin{figure}[!htbp]
\begin{center}
	{\bf {GARCH predicted volatility}}
\includegraphics[height=3.0cm, width=14cm]{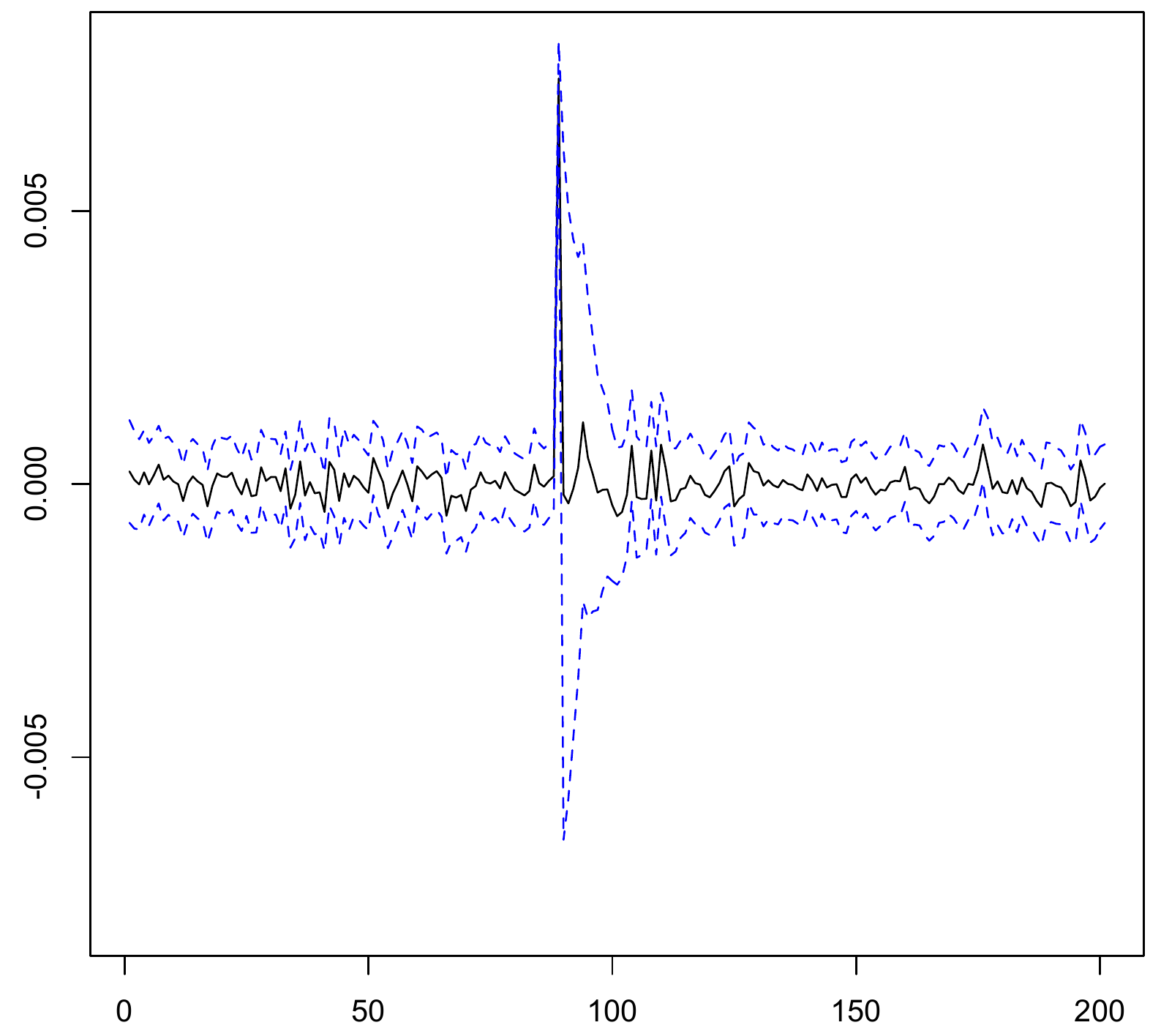}
\end{center}
\caption{One-step-ahead predicted log-volatility, with $\pm 2$ standard prediction errors for two hundred observations from DISCOVER stock exchange.}
\label{figure:fig12}
\end{figure}

\begin{figure}[!htbp]
\begin{center}
	{\bf {GARCH predicted volatility}}
\includegraphics[height=3.0cm, width=14cm]{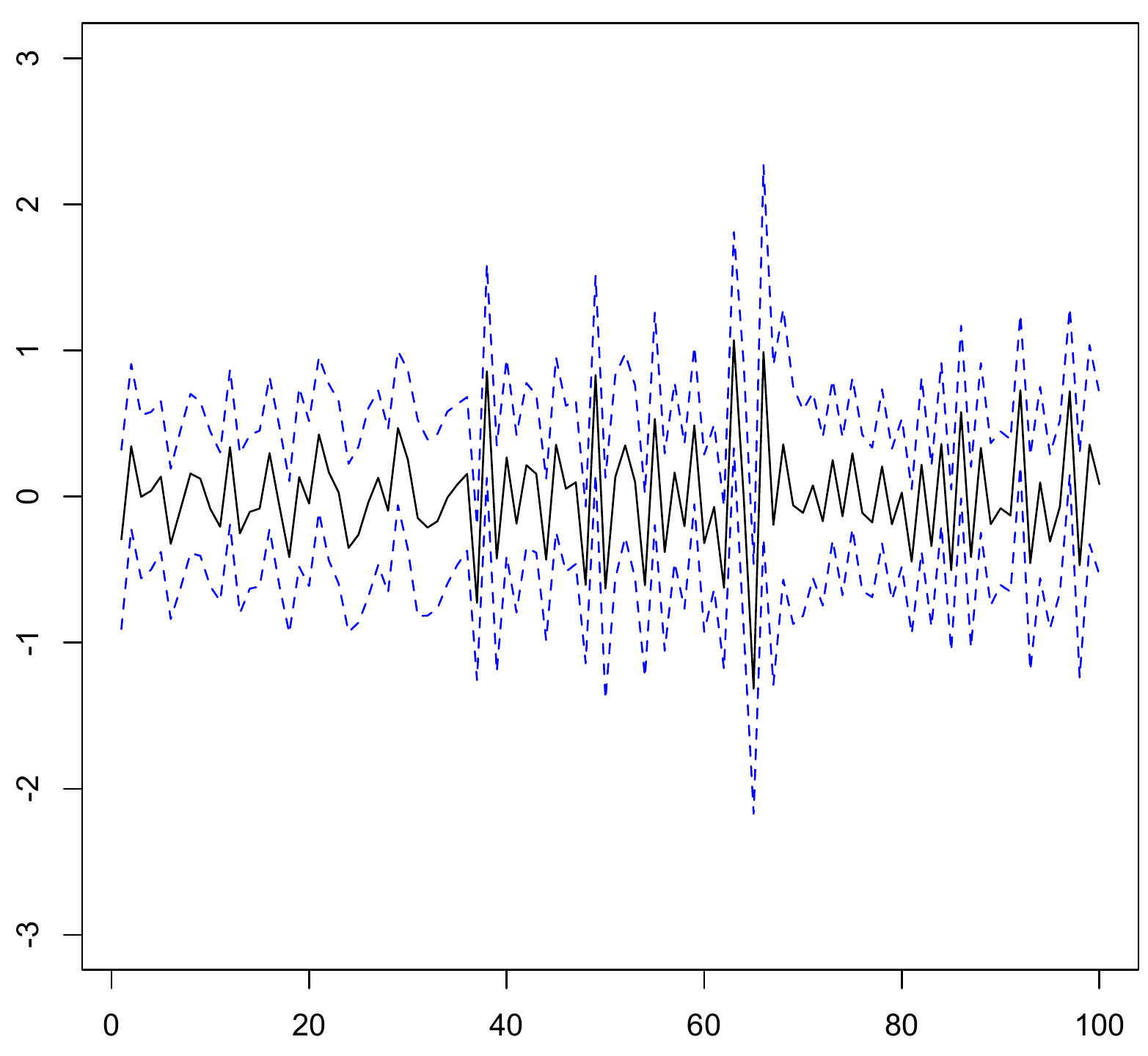}
\end{center}
\caption{One-step-ahead predicted log-volatility, with $\pm 2$ standard prediction errors for one hundred observations from INTEL stock exchange.}
\label{figure:fig13}
\end{figure}

\begin{figure}[!htbp]
\begin{center}
	{\bf {GARCH predicted volatility}}
\includegraphics[height=3.0cm, width=14cm]{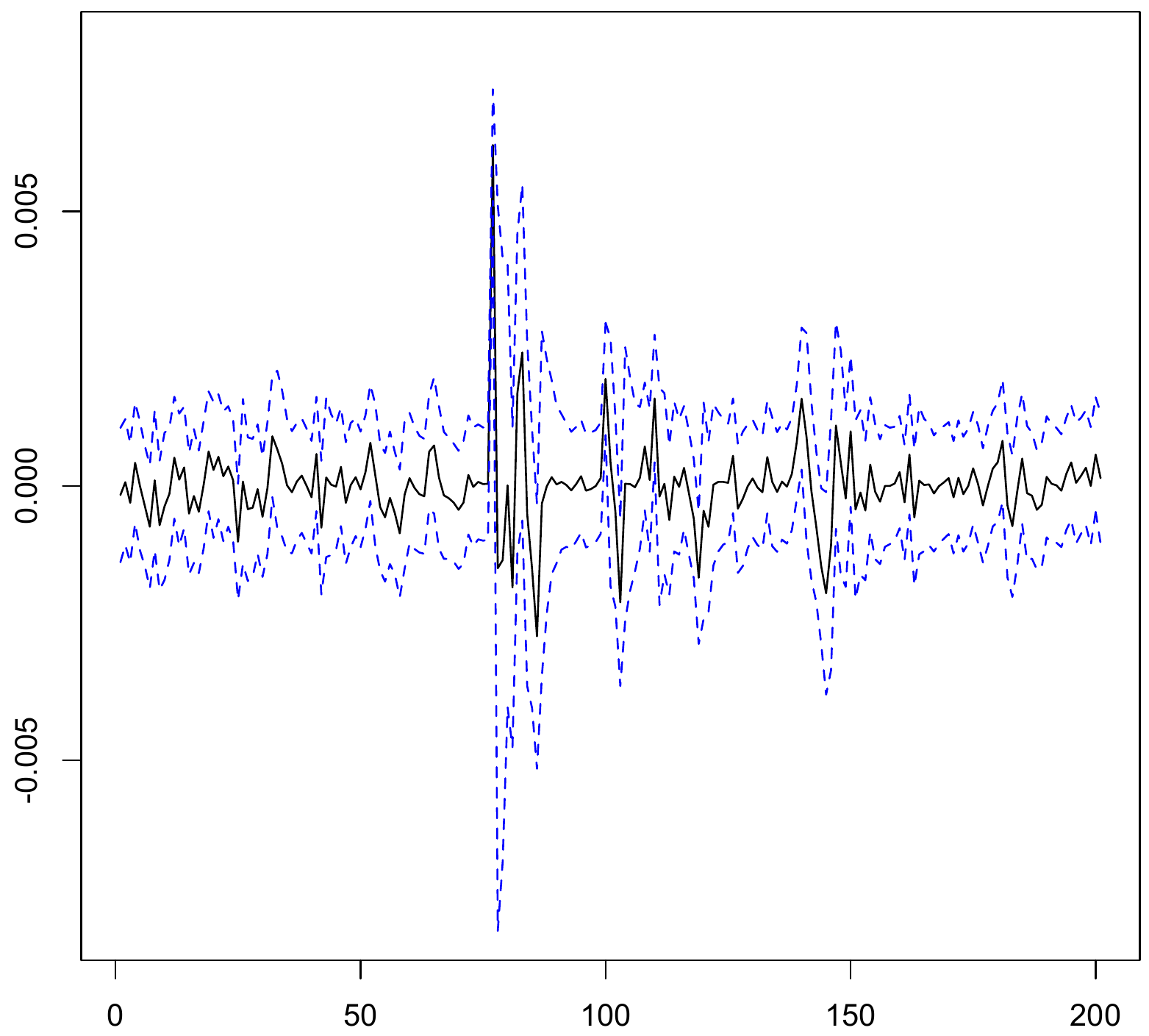}
\end{center}
\caption{One-step-ahead predicted log-volatility, with $\pm 2$ standard prediction errors for two hundred observations from IAG stock exchange.}
\label{figure:fig14}
\end{figure}

To facilitate the understanding of forecasting concepts, we superimpose the plot of one-step-ahead predicted volatility and $\pm 2$ standard prediction errors in Figs. \ref{figure:fig11}-\ref{figure:fig14}. The predicted volatility with $\pm 2 \widehat{\sigma_t}$ is displayed as a dashed line surrounding the original output. It visually shows how values of predicted volatility differ over time.

We also see some limitations of the GARCH (1,1) model itself, when applied to our financial data sets. The positive and negative high frequency returns have the same effect, because volatility depends on squared returns. Thus it does not help to understand the source of variations of a financial time series, i.e. the causes of the variation in volatility. The model provides only a mechanical way to describe the behavior of conditional variance. The blue dashed lines during the financial crash (in Figs. \ref{figure:fig11}-\ref{figure:fig14}) indicate that the model tends to arbitrarily predict volatility, because it slowly responds to large isolated returns. 
\begin{table}[!htbp]
\caption{\bf{GARCH statistics for BAC stock exchange}} 
\centering 
\begin{tabular}{c c c c c} 
\hline\hline 
Parameter & Estimate & Error & t-statistic & p-value  \\ [0.5ex] 
		\hline
		$a_0$  & 1.25E-07 &  1.42E-08 &   8.788 & $<$2E-06 \\ 
		$a_1$ & 2.47E-01 & 5.91E-02 &   4.171 & 3.0E-05\\
		$b_1$ & 7.21E-01 & 3.63E-02  &  19.84 &  $<$2E-16\\		
\hline \hline
\end{tabular}
	\label{table:7} 
\end{table}
\begin{table}[!htbp]
\caption{\bf{GARCH statistics for the DISCOVER stock exchange}} 
\centering 
\begin{tabular}{c c c c c} 
\hline\hline 
Parameter & Estimate & Error & t-value & p-value \\ [0.5ex] 
		\hline
	$a_0$ & 2.86E-08 & 3.89E-09 & 7.349 & 2.0E-13 \\
	$a_1$ & 1.79E-01 & 2.42E-02 & 7.401 & 1.4E-13 \\
	$b_1$ & 7.09E-01 & 3.14E-02 & 22.58 & $<$2E-16 \\
\hline  \hline
\end{tabular}
	\label{table:10} 
\end{table}
\begin{table}[!htbp]
\caption{\bf{GARCH statistics for INTEL stock exchange}} 
\centering 
\begin{tabular}{c c c c c} 
\hline\hline 
Parameter & Estimate & Error & t-statistic & p-value  \\ [0.5ex] 
		\hline
		$a_0$  & 0.02167  & 0.00982  &  2.206 & 0.0274 \\ 
		$a_1$ & 0.15914 & 0.03682  &  4.322 & 1.5E-05\\
		$b_1$   & 0.61185 & 0.13686 & 4.471 & $<$7.8E-06\\		
\hline \hline 
\end{tabular}
	\label{table:8} 
\end{table}
\begin{table}[!htbp]
\caption{\bf{GARCH statistics for IAG stock exchange}} 
\centering 
\begin{tabular}{c c c c c} 
\hline\hline 
Parameter & Estimate & Error & t-statistic & p-value  \\ [0.5ex] 
		\hline
	$a_0$ & 1.31E-07 & 1.59E-08 & 8.278 & 2.2E-16 \\
	$a_1$ & 2.77E-01 &  4.03E-02  & 6.867 & 6.6E-12\\
	$b_1$ & 4.83E-01  & 5.24E-02  & 9.208 &  $<$2E-16 \\
\hline \hline 
\end{tabular}
	\label{table:9} 
\end{table}
\begin{table}[!htbp]
	\caption{\bf{Standardised Residuals Tests for BAC stock exchange}} 
	\centering 
	\begin{tabular}{c c c c c} 
		\hline\hline 
		 & Residuals &  Tests & Statistics & p-value  \\ [0.5ex] 
		\hline
		Jarque-Bera Test  & $R$   & $\chi^2$ &  4918340 & 0  \\ 
		Shapiro-Wilk Test   & $R$  &  W   &   0.649712 & 0         \\
	    Ljung-Box Test  & $R$  &  Q(10) & 14.71311 & 0.1429 \\
		Ljung-Box Test  &  $R$  &  Q(15) & 17.82156 & 0.2722\\
		Ljung-Box Test   & $R$ & Q(20) & 18.77450 &  0.5365 \\
		Ljung-Box Test   & $R^2$ & Q(10) & 0.107465 & 1\\
		Ljung-Box Test & $R^2$ &  Q(15) &  0.130187 & 1     \\
		Ljung-Box Test  & $R^2$ &  Q(20) &  0.147051 & 1  \\
		LM-Arch Test  & $R$  &  T$R^2$ &  0.126933 & 1      \\    
		\hline \hline
	\end{tabular}
	\label{table:11} 
\end{table}
\begin{table}[!htbp]
	\caption{\bf{Standardised Residuals Tests for DISCOVER stock exchange}} 
	\centering 
	\begin{tabular}{c c c c c} 
		\hline\hline 
		 &  &  & Statistics & P-value  \\ [0.5ex] 
		\hline
		Jarque-Bera Test  & R   & $\chi^2$ &  1688779 &   0  \\ 
		Shapiro-Wilk Test   & R   &  W   &   0.722888 & 0         \\
	    Ljung-Box Test    & R  &  Q(10) & 10.93070  & 0.3629 \\
		Ljung-Box Test   &  R  &  Q(15) & 13.58973 & 0.5568 \\
		Ljung-Box Test   &  R  &  Q(20) & 16.81071 & 0.6652 \\
		Ljung-Box Test   & $R^2$ & Q(10) & 0.198119 & 0.9999 \\
		Ljung-Box Test   & $R^2$ & Q(15) & 0.258743 & 1 \\
		Ljung-Box Test & $R^2$ &  Q(20) &  0.369189 & 1 \\
		LM Arch Test & R  &  $TR^2$ &   0.208586 & 1 \\    
		\hline \hline
	\end{tabular}
	\label{table:12} 
\end{table}
\begin{table}[!htbp]
	\caption{\bf{Standardised Residuals Tests for INTEL stock exchange}} 
	\centering 
	\begin{tabular}{c c c c c} 
		\hline\hline 
		 & Residuals &  Tests & Statistics & p-value  \\ [0.5ex] 
		\hline
		Jarque-Bera Test  & $R$   & $\chi^2$ & 1769.854 &   0  \\ 
		Shapiro-Wilk Test   & $R$   &  W   &   0.971157 & 0         \\
	    Ljung-Box Test    & $R$  &  Q(10) & 296.0897 & 0\\
		Ljung-Box Test   &  $R$  &  Q(15) & 301.6321 & 0\\
		Ljung-Box Test   & $R$ & Q(20) & 304.7640 & 0 \\
		Ljung-Box Test   & $R^2$ & Q(10) & 7.968585 & 0.6319\\
		Ljung-Box Test & $R^2$ &  Q(15) &  16.48274 & 0.3507\\
		Ljung-Box Test & $R^2$ &  Q(20) &  59.49665 & 8.5E-06  \\
		LM-Arch Test & R  &  T$R^2$ & 14.76281 & 0.2547 \\    
		\hline  \hline
	\end{tabular}
	\label{table:13} 
\end{table}
\begin{table}[!htbp]
	\caption{\bf{Standardised Residuals Tests for IAG stock exchange}} 
	\centering 
	\begin{tabular}{c c c c c} 
		\hline\hline 
		 & Residuals &  Tests & Statistics & p-value  \\ [0.5ex] 
		\hline
		Jarque-Bera Test  & $R$   & $\chi^2$ & 126041.2 &   0  \\ 
		Shapiro-Wilk Test   & $R$   &  W   &   0.842114 & 0    \\
	    Ljung-Box Test    & $R$  &  Q(10) & 12.66334 & 0.2431 \\
		Ljung-Box Test   &  $R$  &  Q(15) &  15.02200 & 0.4498 \\
		Ljung-Box Test   & $R$ & Q(20) & 21.73719 & 0.3549 \\
		Ljung-Box Test & $R^2$&  Q(10) &  0.146992 & 1 \\
		Ljung-Box Test & $R^2$ &  Q(15) & 0.333308 & 1 \\
		Ljung-Box Test & $R^2$ &  Q(20) & 0.706543 &  1\\
		LM-Arch Test  & R  &  T$R^2$ & 0.243845 &  1 \\    
		\hline \hline
	\end{tabular}
	\label{table:14} 
\end{table}
\newpage
\section{Analysis of Stochastic Model}
\label{sec6}
In this section we use the stochastic volatility model to forecast the volatility of a time series. The first subsection is devoted to analyzing high frequency financial time series and the latter subsection includes geophysical time series.

\begin{figure}[!htbp]
    \centering
    \subfigure[Earthquake (TUC station)]
    { %
        \includegraphics[height=2.3cm,width=5cm]{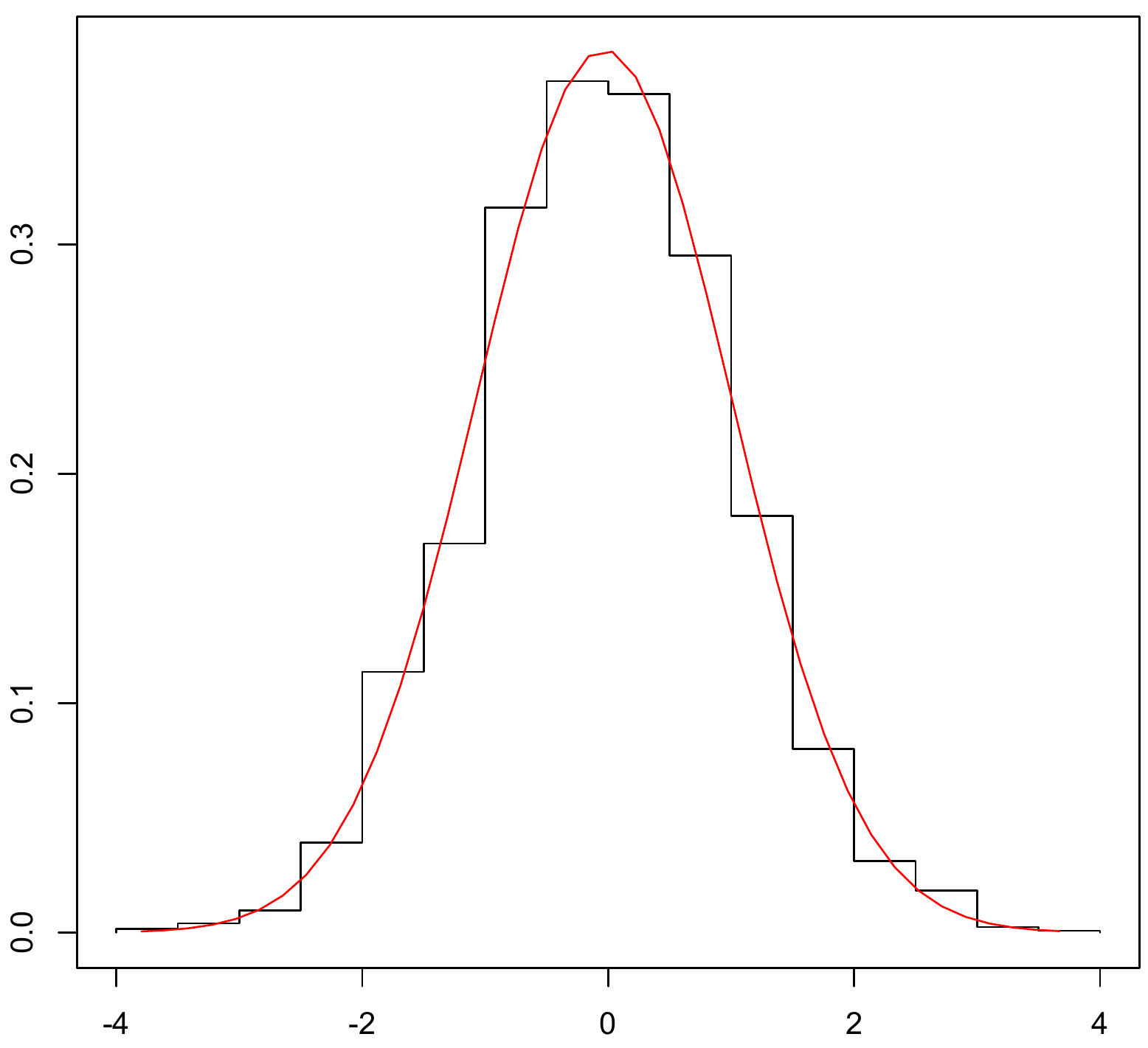}
    }
    \quad
  \subfigure[Explosion (TUC station)]
    {  
        \includegraphics[height=2.3cm,width=5cm]{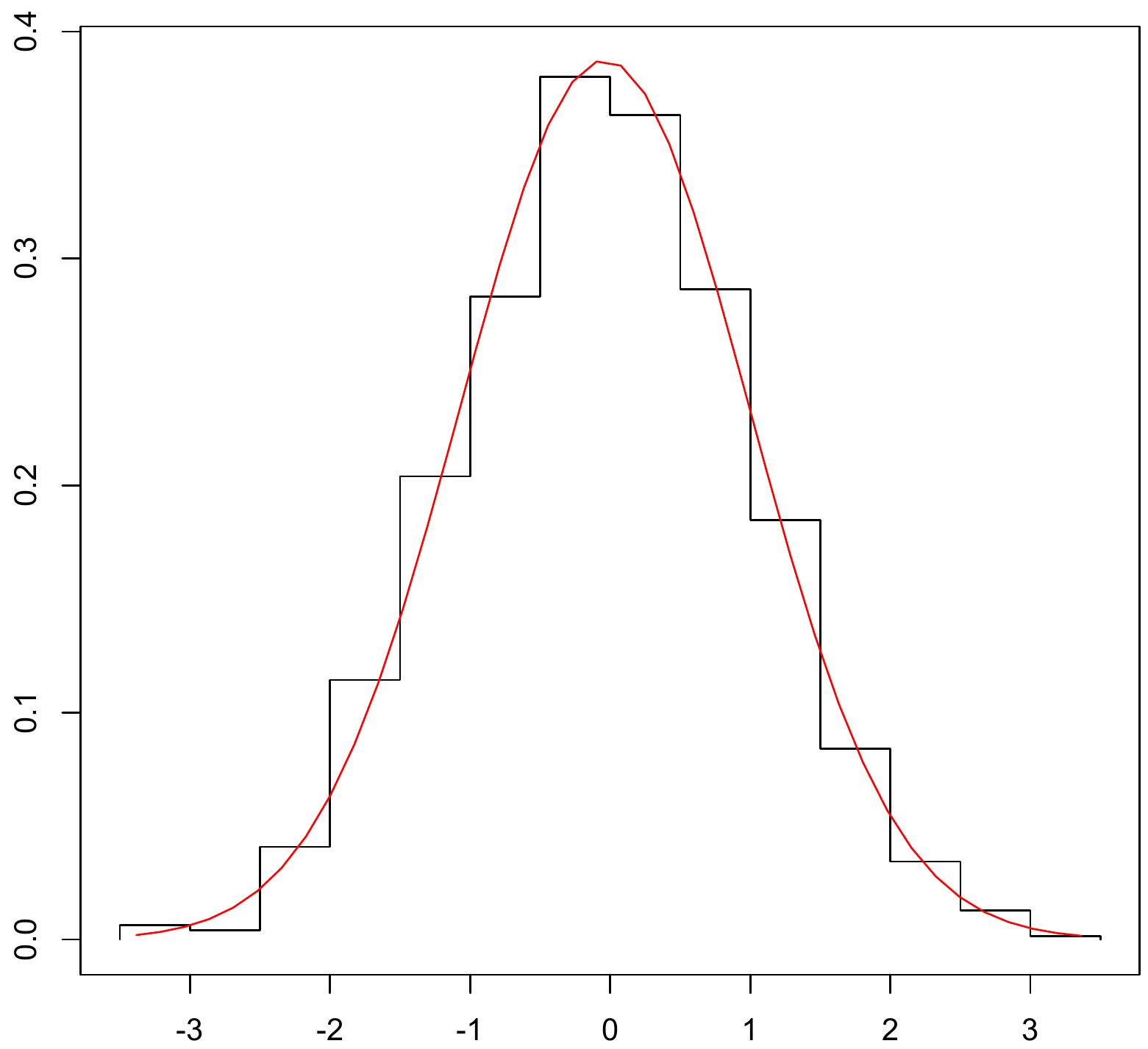}
    }
   \linebreak

    \subfigure[Earthquake (ANMO station)]
    {
        \includegraphics[height=2.3cm,width=5cm]{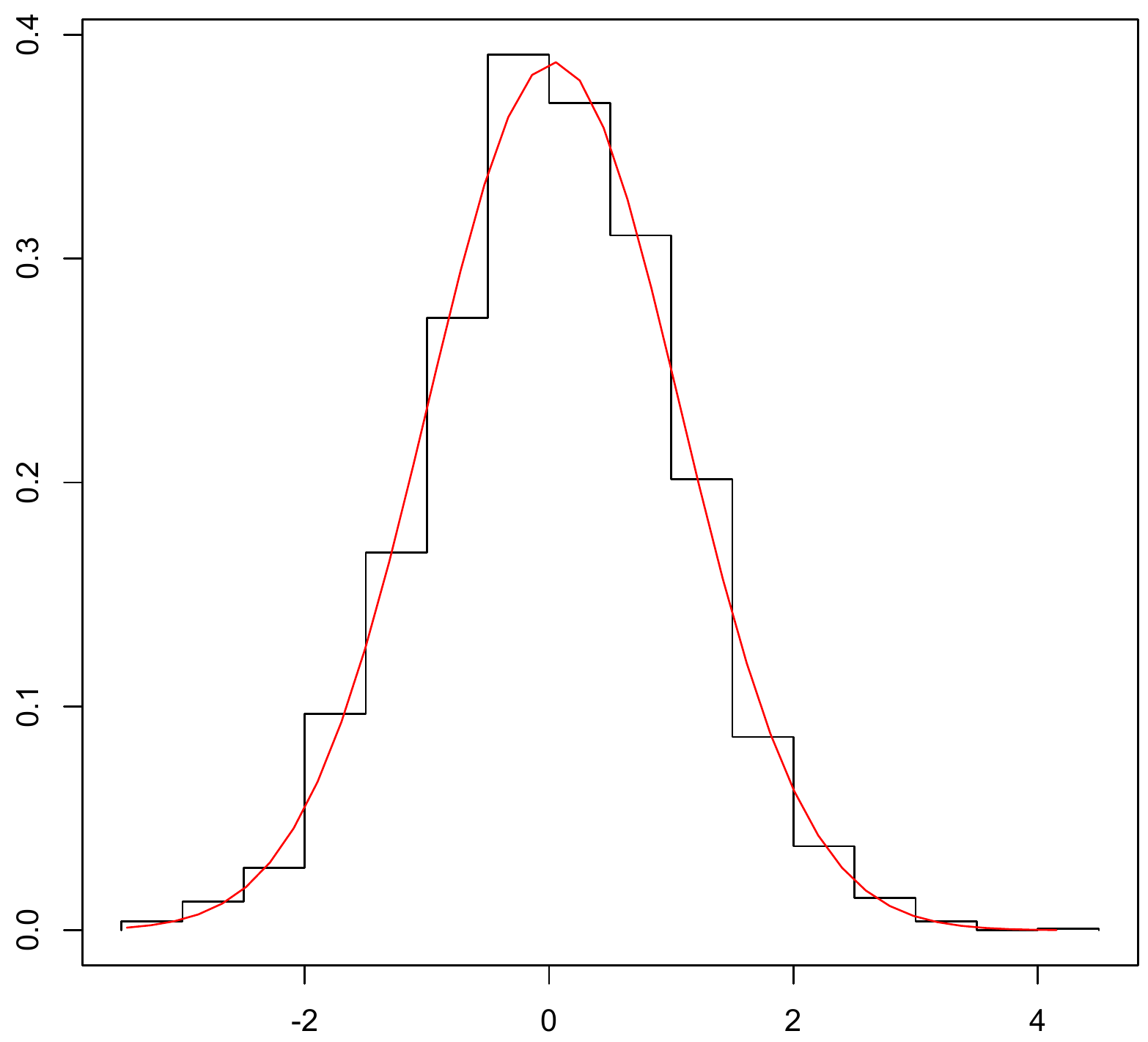}
    }
    \quad
    \subfigure[Explosion (ANMO station)]
        {
            \includegraphics[height=2.3cm,width=5cm]{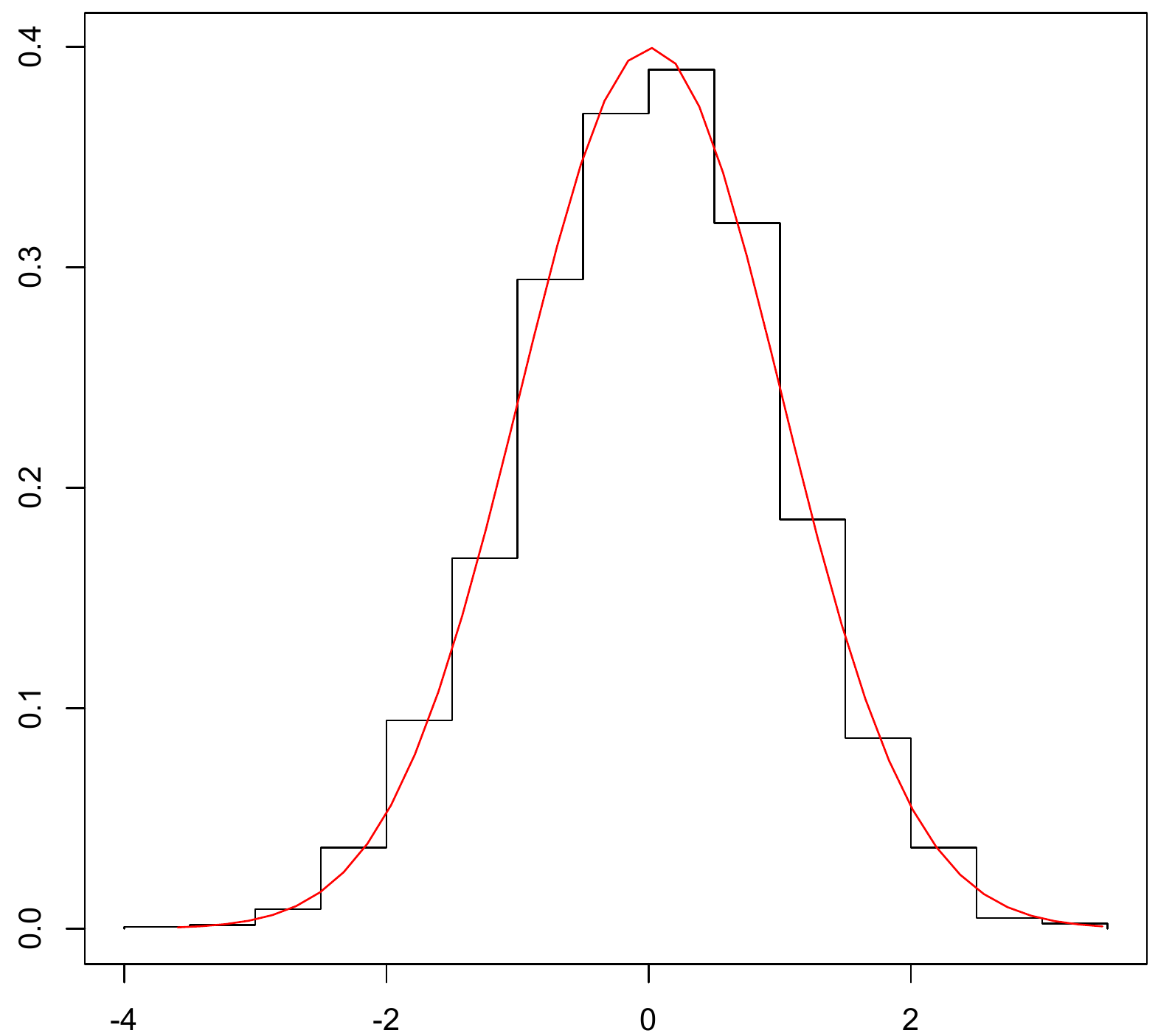}
        }
       	\caption{The histograms of geophysical time series and the fitted Normal density.}
          \label{figure:fig15}
\end{figure}

The GARCH model used in the previous section differs from the SV model because it does not have any stochastic noise. The SV model is characterized by the fact that it invariably contains its probability density function. We compute the maximum likelihood by taking into consideration the conditional Normal distribution. From Figs. \ref{figure:fig10} and \ref{figure:fig15} we conclude that the histograms of the financial and geophysical time series respectively, are well represented. 

The parameters from time-varying Eqs. (\ref{eq:11}) and (\ref{eq:12}) were initialized in order to observe the performance of the SV algorithms during a set of magnitudes for each seismic event. We set the initial values to be
$\alpha_0=0,\alpha_1=0.96,\sigma_\omega=0.3,\sigma_0=1,\phi_1=-4,\sigma_1=3$ and $\lambda$, the mean of the observations. In order to maximize  Eq. (\ref{eq:21}), the innovation processes for Eqs. (\ref{eq:11}) and (\ref{eq:12}) were fitted to the data by taking into consideration this time-varying probability ($p_1=0.5$).

\subsection{Volatility of Financial Time Series}
Using the stochastic volatility model, we estimate the parameters and one-step-ahead predicted log-volatility of high frequency returns from four stock exchanges.

\begin{table}[!htbp]
\caption{\bf{Summary statistics for DISCOVER stock exchange}} 
\centering 
\begin{tabular}{c c c } 
\hline\hline 
Parameter & Estimate & Standard Error \\ [0.5ex] 
		\hline
$\alpha_0$  & 0.0053 & 0.0008
\\
$\alpha_1$  &  0.9799 & 0.0036
\\
$\sigma_{\omega}$  & 0.1841 & 0.0004
\\
$\lambda$ &-16.801 & 0.1687
\\
$\sigma_0$  & 1.0697 & 0.0008

\\
$\phi_1$    &-2.2973 & 0.0011
\\
$\sigma_1$    & 2.8771 & 0.0008

\\

\hline \hline 
\end{tabular}
	\label{table:93} 
\end{table}

\begin{table}[!htbp]
\caption{\bf{Summary statistics for INTEL stock exchange}} 
\centering 
\begin{tabular}{c c c } 
\hline\hline 
Parameter & Estimate & Standard Error \\ [0.5ex] 
		\hline
$\alpha_0$  & -0.1650 & 0.4123

\\
$\alpha_1$  &  0.8629 & 0.0682

\\
$\sigma_{\omega}$     & 0.5978 & 0.1871
\\
$\lambda$ &-14.166 & 2.9146

\\
$\sigma_0$  & 0.9478 & 0.1311

\\
$\phi_1$    &-3.3405 & 0.1904
\\
$\sigma_1$    & 3.2604 & 0.1158

\\

\hline \hline 
\end{tabular}
	\label{table:94} 
\end{table}

\begin{table}[h!]
\caption{\bf{Summary statistics for IAG stock exchange}} 
\centering 
\begin{tabular}{c c c } 
\hline\hline 
Parameter & Estimate & Standard Error \\ [0.5ex] 
		\hline
$\alpha_0$  & -0.1994 & 0.4848\\
$\alpha_1$  &  0.8440 &0.0738\\
$\sigma_{\omega}$     & 0.6462 &0.1976\\
$\lambda$ &-14.164 & 3.0113\\
$\sigma_0$  & 0.9536 & 0.1434\\
$\phi_1$    &-3.3967 & 0.2038\\
$\sigma_1$    & 3.6077 & 0.1261\\

\hline \hline 
\end{tabular}
	\label{table:91} 
\end{table}

\begin{table}[!htbp]
\caption{\bf{Summary statistics for BAC stock exchange}} 
\centering 
\begin{tabular}{c c c } 
\hline\hline 
Parameter & Estimate & Standard Error \\ [0.5ex] 
		\hline
$\alpha_0$  & -0.1596 & 0.4066\\
$\alpha_1$  &  0.8641 &0.0676
\\
$\sigma_{\omega}$     & 0.5949 & 0.1856\\
$\lambda$ &-14.191 & 2.9078
\\
$\sigma_0$  & 0.9471 & 0.1299
\\
$\phi_1$    &-3.3374 & 0.1895\\
$\sigma_1$    &3.238 & 0.1152
\\

\hline \hline 
\end{tabular}
	\label{table:92} 
\end{table}


\begin{figure}[!htbp]
	\begin{center}
		{\bf {Predicted log-volatility of DISCOVER stock exchange}}
		\includegraphics[height=3.5cm, width=14cm]{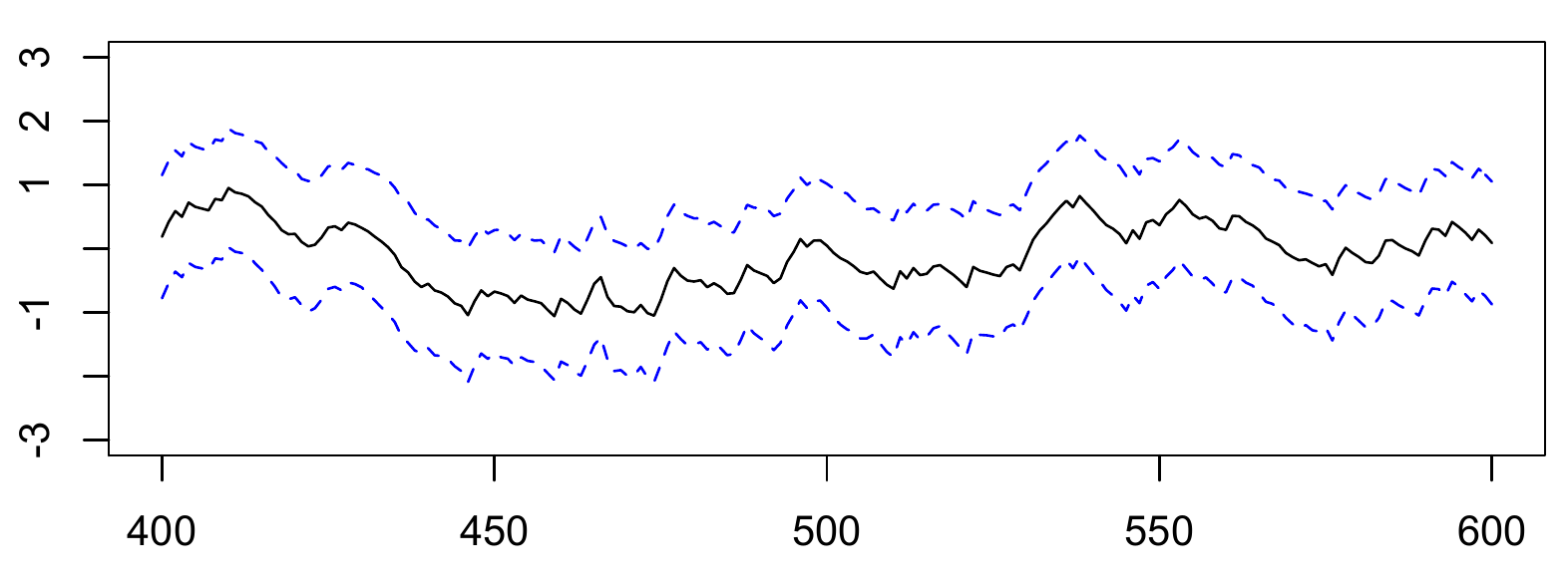}
	\end{center}
	\caption{One-step-ahead predicted log-volatility, with $\pm 2$ standard prediction errors for DISCOVER stock exchange.}
	\label{figure:fig116}
\end{figure}

\begin{figure}[!htbp]
	\begin{center}
		{\bf {Predicted log-volatility of INTEL stock exchange}}
		\includegraphics[height=3.5cm, width=14cm]{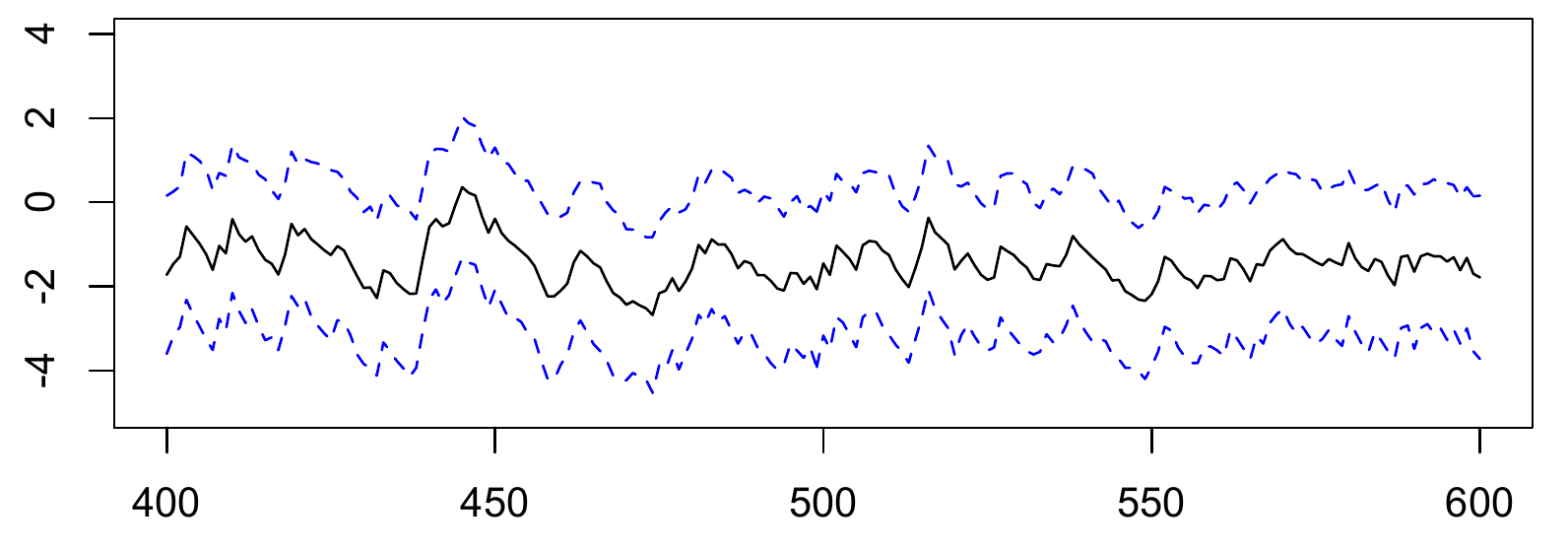}
	\end{center}
	\caption{One-step-ahead predicted log-volatility, with $\pm 2$ standard prediction errors for INTEL stock exchange.}
	\label{figure:fig126}
\end{figure}


\begin{figure}[h!]
	\begin{center}
		{\bf {Predicted log-volatility of IAG stock exchange}}
		\includegraphics[height=3.5cm, width=14cm]{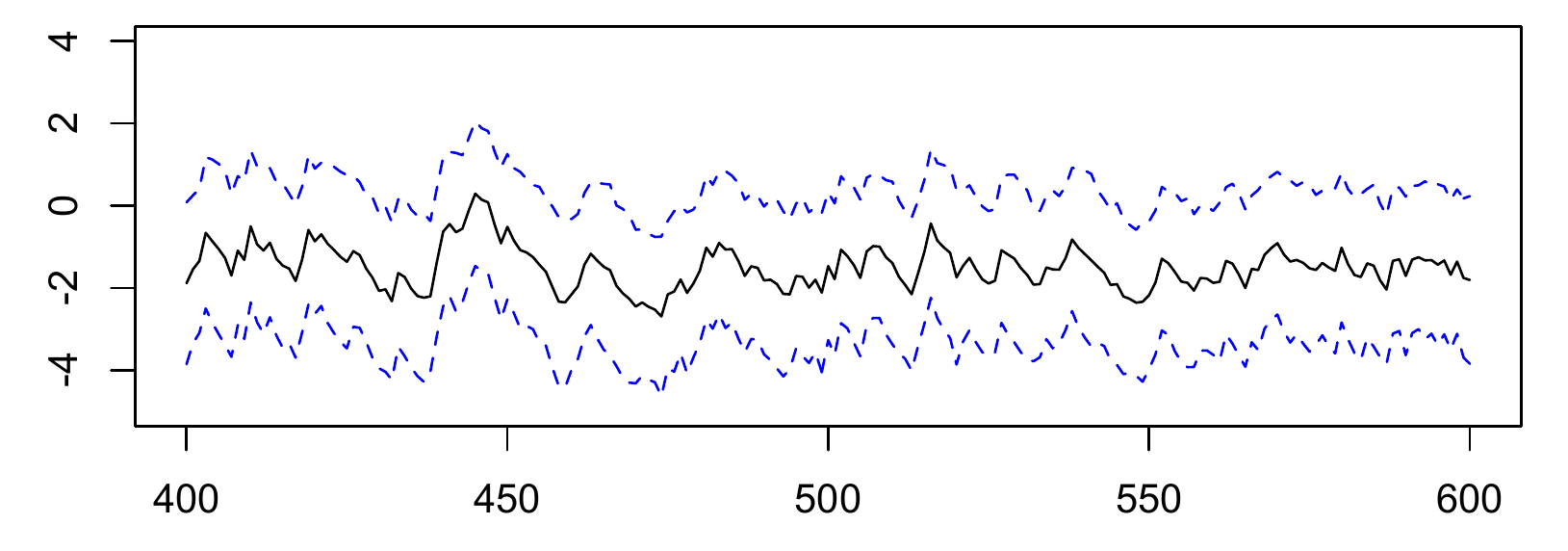}
	\end{center}
	\caption{One-step-ahead predicted log-volatility, with $\pm 2$ standard prediction errors for IAG stock exchange.}
	\label{figure:fig136}
\end{figure}

\begin{figure}[!htbp]
	\begin{center}
		{\bf {Predicted log-volatility of BAC stock exchange}}
		\includegraphics[height=3.5cm, width=14cm]{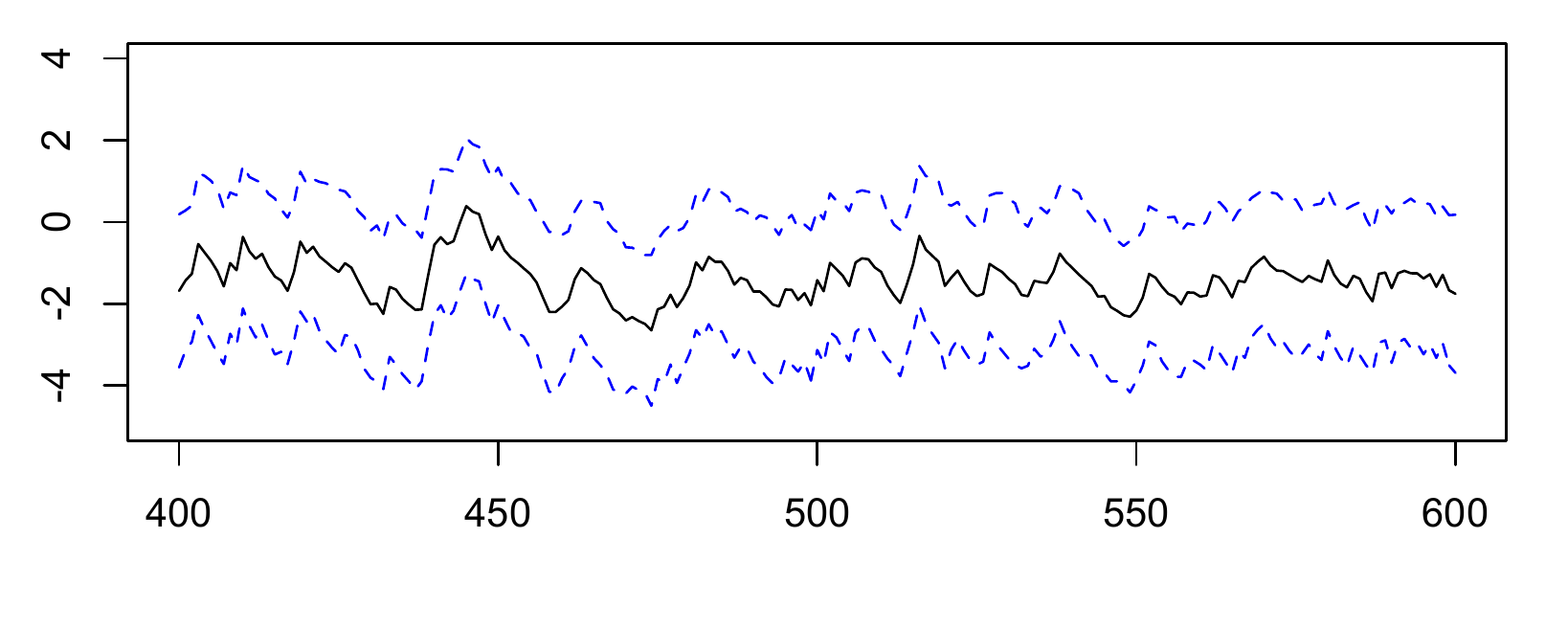}
	\end{center}
	\caption{One-step-ahead predicted log-volatility, with $\pm 2$ standard prediction errors for BAC stock exchange.}
	\label{figure:fig106}
\end{figure}






\subsection{Volatility of Geophysical Time Series}

Similarly, using the stochastic volatility model we estimate the parameters and one-step-ahead predicted log-volatility of earthquake and explosion time series.

\begin{table}[!htbp]
	\caption{\bf{Summary statistics for Earthquake data sets}} 
	\centering 
\begin{tabular}{|c |c     c | c    c |c l}
\cline{1-5}
\multicolumn{1}{ |c|  }{\multirow{2}{*}{Parameter} } &
 \multicolumn{2}{|c|}{TUC station} &        \multicolumn{2}{|c|}{ANMO station}    \\ \cline{2-5}
\multicolumn{1}{ |c|  }{}                        &
 Estimates &    Standard Error  & Estimates &    Standard Error    &     \\ \cline{1-5}
		$\alpha_0$ & 0.0286 & 0.0343   & 0.0828 & 0.1015\\ 
		$\alpha_1$ & 0.9960 & 0.0013 & 0.9851 & 0.0028\\
		$\sigma_\omega$ & 0.4534 & 0.0357  & 0.7427 & 0.0183 \\
		$\lambda$ & -4.0974 & 7.4020    &  -5.2073 & 6.4241 \\
    	$\sigma_0$ & 0.6284 & 0.0401 & 0.0007 & 0.0720 \\
		$\phi_1$ & -2.4730 & 0.0862 & -2.3090 & 0.0773  \\
	   $\sigma_1$ & 2.3481 & 0.0518  & 2.1530 & 0.0482  \\ [1ex] 
	\cline{1-5} 
	\end{tabular}
	\label{table:15} 
\end{table}
\begin{table}[!htbp]
	\caption{\bf{Summary statistics for Explosion data sets}} 
	\centering 
\begin{tabular}{|c |c     c | c    c |c l}
\cline{1-5}
\multicolumn{1}{ |c|  }{\multirow{2}{*}{Parameter} } &
 \multicolumn{2}{|c|}{TUC station} &        \multicolumn{2}{|c|}{ANMO station}    \\ \cline{2-5}
\multicolumn{1}{ |c|  }{}                        &
 Estimates &    Standard Error  & Estimates &    Standard Error    &     \\ \cline{1-5}%
		$\alpha_0$ & 0.1778 & 0.1025 & 0.1889 & 0.1121 \\ 
		$\alpha_1$ & 0.9874 & 0.0028 &  0.9814 & 0.0032  \\
		$\sigma_\omega$ & 0.7492  & 0.0181 & 0.7003 & 0.0171 \\
		$\lambda$ & -10.151 & 6.7569  & -10.076 & 5.2518  \\
		$\sigma_0$ & 2.13E-05 & 0.0517  & 7.09E-05 & 0.0501 \\
		$\phi_1$ & -2.3475 & 0.0798 & -2.3063 & 0.0740 \\
		$\sigma_1$ & 2.1657 & 0.0491 & 2.0917& 0.0462 \\ [1ex] 
		\cline{1-5} 
	\end{tabular}
	\label{table:16} 
\end{table}
\begin{figure}[!htbp]
	\begin{center}
		{\bf {Predicted log-volatility of Earthquake data}}
		\includegraphics[height=3.5cm, width=14cm]{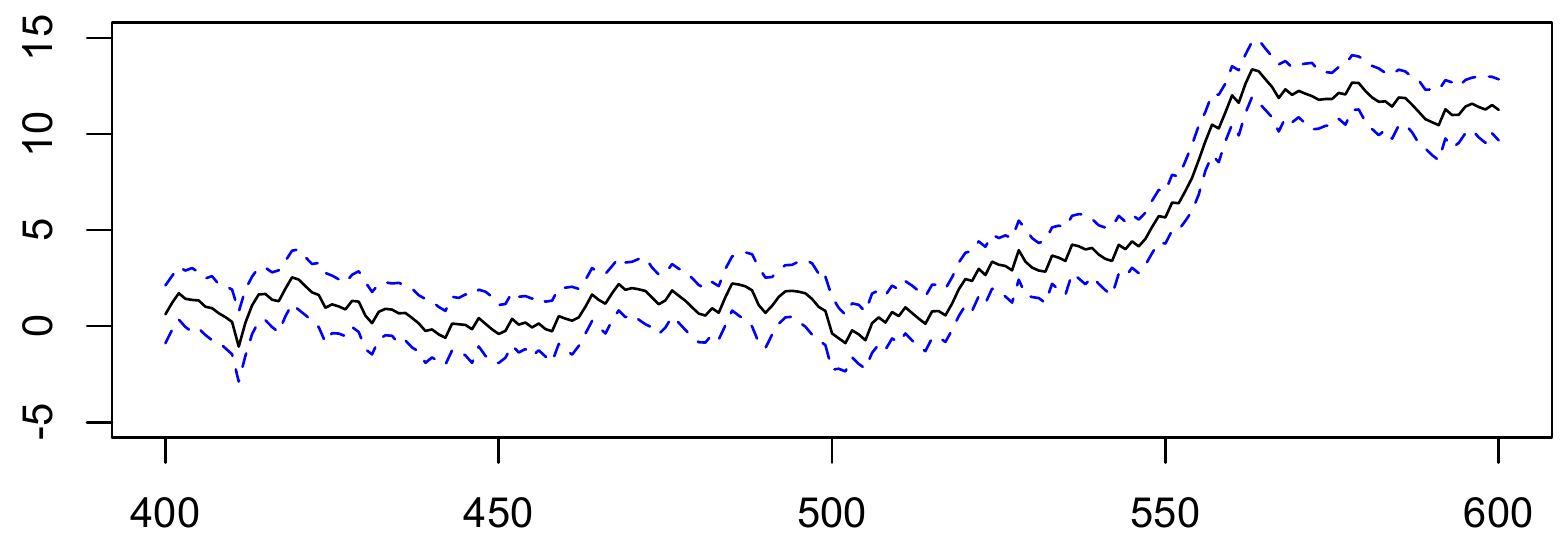}
	\end{center}
	\caption{One-step-ahead predicted log-volatility, with $\pm 2$ standard prediction errors for two hundred observations from TUC station.}
	\label{figure:fig16}
\end{figure}
\begin{figure}[!htbp]
	\begin{center}
		{\bf {Predicted log-volatility of Explosion data}}
		\includegraphics[height=3.5cm, width=14cm]{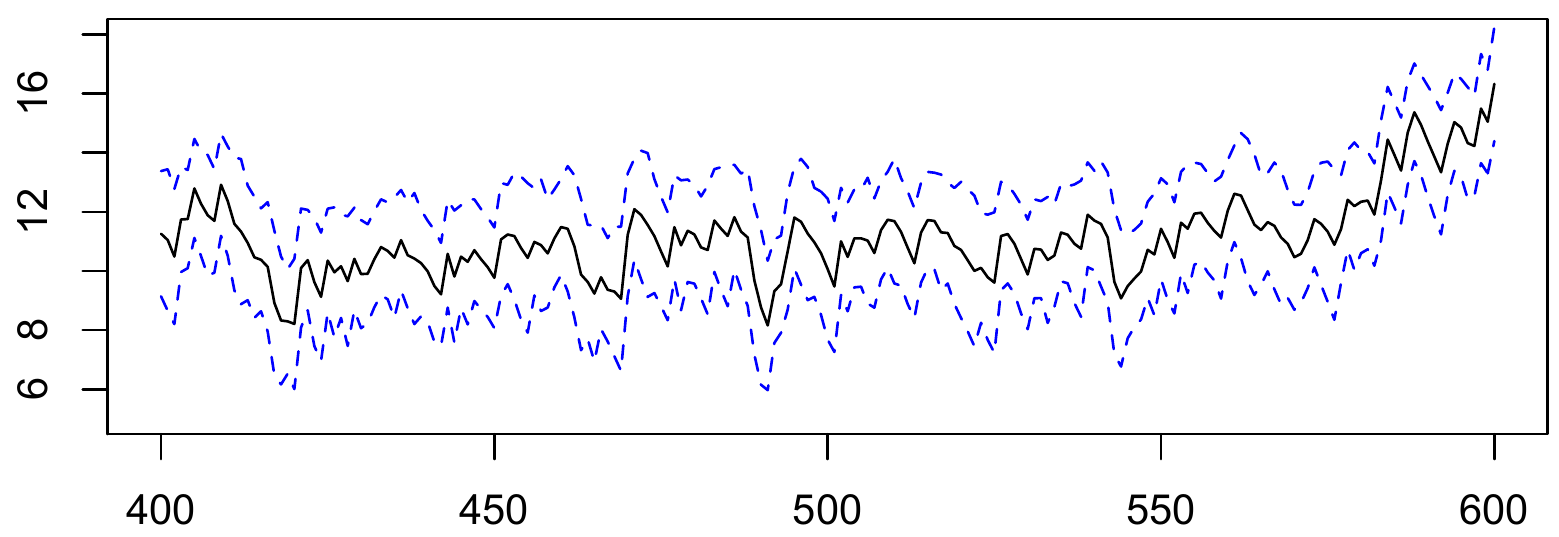}
	\end{center}
	\caption{One-step-ahead predicted log-volatility, with $\pm 2$ standard prediction errors for two hundred observations from TUC station.}
	\label{figure:fig17}
\end{figure}
\begin{figure}[!htbp]
	\begin{center}
		{\bf {Predicted log-volatility of Earthquake data}}
		\includegraphics[height=3.5cm, width=14cm]{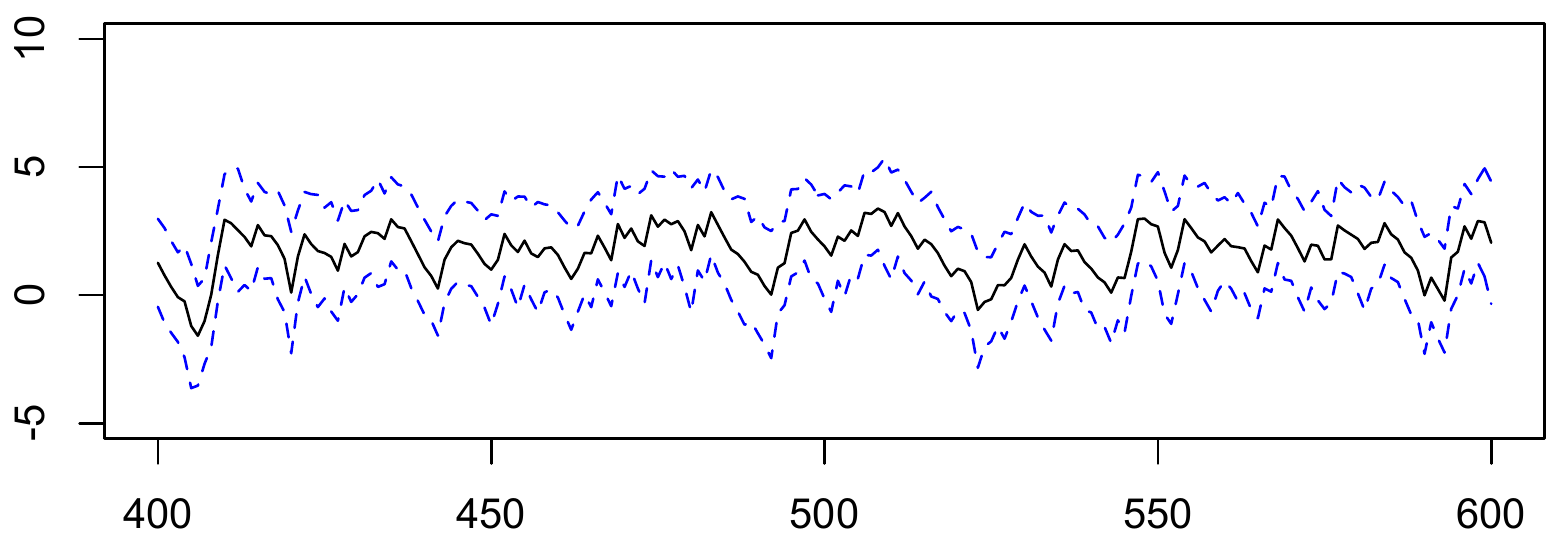}
	\end{center}
	\caption{One-step-ahead predicted log-volatility, with $\pm 2$ standard prediction errors for two hundred observations from ANMO station.}
	\label{figure:fig18}
\end{figure}
\begin{figure}[!htbp]
	\begin{center}
		{\bf {Predicted log-volatility of Explosion data}}
		\includegraphics[height=3.5cm, width=14cm]{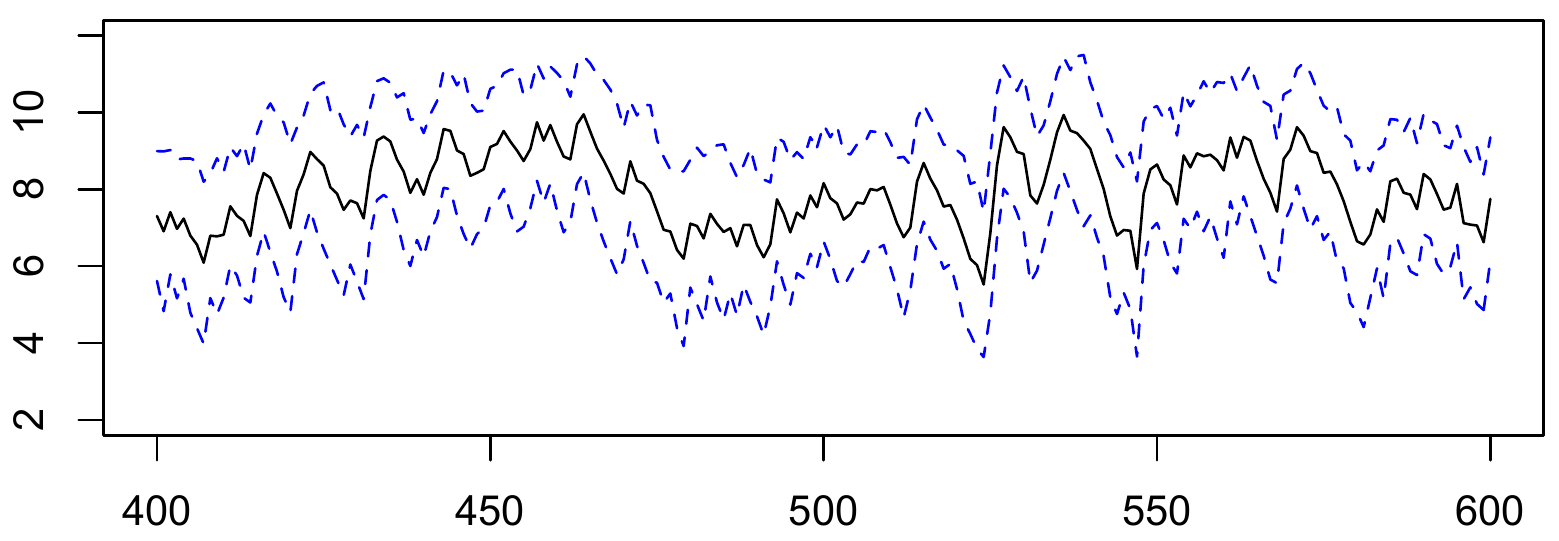}
	\end{center}
	\caption{One-step-ahead predicted log-volatility, with $\pm 2$ standard prediction errors for two hundred observations from ANMO station.}
	\label{figure:fig19}
\end{figure}
\newpage
\subsection{Discussion}

Tables \ref{table:91} to \ref{table:16} summarize the estimation of parameters ($\alpha_0, \alpha_1$, $\sigma_w$, $\lambda$,  $\sigma_0$, $\phi_1$, and  $\sigma_1$). The estimated error in these tables makes two things evident: firstly, the estimates are close to the true parameters; secondly, the algorithm of the SV model is consistent with the results obtained by using the data. The variance $\sigma_w^2$ of the log-volatility process measures the uncertainty related to the future  data volatility. If the value of $\sigma_w^2$ is zero, it is not possible to identify the SV model. The parameter $\alpha_1$  is considered as a measure of the persistence of shocks to the volatility. Tables \ref{table:91} to \ref{table:16} indicate that $\alpha_1$ is less than 1, which suggests that the latent volatility process and $y_t$ are stationary, what confirms the results of section \ref{sec4}.

In these tables, we notice that $\alpha_1$ is near to unity and $\sigma_w^2 $ is different from 0, which means that the volatility evolution is not smooth over time. It also suggests that the time series could be heteroscedastic by nature, that is, there is a non-constant conditional volatility over time. So, it is very useful to control the risk or to mitigate the effect of hazards. For example, if there are two time series having the same mean but with different variances, we would then consider the series with lower variance, because it is less risky. 

\section{Conclusion}
\label{sec7}
In this study, we have used high frequency financial returns and geophysical time series  measured  every minute. We have implemented a type of volatility models that incorporates time-varying parameters in a stationary scenario. To obtain a good fit for the data, we used a deterministic model for high-frequency financial returns and a stochastic model for the financial and geophysical time series. The geophysical data aligns with the SV model because of its stochastic behavior whereas the GARCH model does not fit the geophysical data. This is because the GARCH model tends to arbitrarily predict volatility in the case of large arrival phases of earthquake and explosion time series. 

The parameter estimates of the GARCH (1,1) model indicate that there exists a stationary solution in the conditional volatility of high frequency financial returns (see Section \ref{sec5}).  For the SV model, we estimated the volatility parameters of time series based on the geophysical and financial time series. The fitted SV model allows us to capture the volatility evolution that suggests the physical and long-memory behavior of the data. With the use of maximum likelihood computation, we have succeeded in making a good prediction despite the variation of the observational noise from a Normal mixture distribution, because the geophysical time series studied is not fully Gaussian (see the histograms in Fig. \ref{figure:fig15}).

Our results suggest that the stochastic process to forecast the time series is more effective in enforcing the characteristic of time-varying parameters than the commonly used deterministic process. It is because the one-step-ahead predictions along with the estimated standard error of stochastic volatility model do not show any limitations unlike the GARCH model (see section \ref{sec5}). We notice that the GARCH model is more sensitive to noise or unexpected shocks than the stochastic volatility model. The advantage of the stochastic methodology is that the estimates obtained are stable around the true value. Moreover, the low errors (see Tables \ref{table:15} and \ref{table:16}) imply that the estimation procedure is accurate, meaning that it is capable of generating a higher forecasting accuracy.
\bigskip

\end{document}